\documentclass[3p,review]{elsarticle}

\usepackage{amsmath,amssymb,amsfonts}
\usepackage{graphicx}
\usepackage{hyperref}
\usepackage{xcolor}
\usepackage{ulem}
\usepackage{amsmath,amsfonts,amssymb,bm,cancel}
\usepackage{soul,cancel}
\usepackage{epsf,amsmath,amssymb,verbatim,color,multirow,pifont,graphicx,cleveref,tabularx}

\journal{Chaos, Solitons \& Fractals}
\begin{document}\
\begin{frontmatter}
\title{$(k,q)$-core decomposition of hypergraphs}

\author[CCSS]{Jongshin Lee}

\author[KU]{Kwang-Il Goh}
\ead{kgoh@korea.ac.kr}

\author[KIAS]{Deok-Sun Lee}
\ead{deoksunlee@kias.ac.kr}

\author[CCSS]{B. Kahng}
\ead{bkahng@kentech.ac.kr}

\affiliation[CCSS] {
    organization={Center for Complex Systems, and KI for Grid Modernization, Korea Institute of Energy Technology},
    city={Naju},
    postcode={58217},
    country={Korea}
}

\affiliation[KU] {
    organization={Department of Physics, Korea University, Korea University},
    city={Seoul},
    postcode={02841},
    country={Korea}
}

\affiliation[KIAS] {
    organization={School of Computational Sciences, and Center for AI and Natural Sciences, Korea Institute for Advanced Study},
    city={Seoul},
    postcode={02455},
    country={Korea}
}

\begin{abstract}
In complex networks, many elements interact with each other in different ways. A hypergraph is a network in which group interactions occur among more than two elements. In this study, first, we propose a method to identify influential subgroups in hypergraphs, named $(k,q)$-core decomposition. The $(k,q)$-core is defined as the maximal subgraph in which each vertex has at least $k$ hypergraph degrees \textit{and} each hyperedge contains at least $q$ vertices. The method contains a repeated pruning process until reaching the $(k,q)$-core, which shares similarities with a widely used $k$-core decomposition technique in a graph. Second, we analyze the pruning dynamics and the percolation transition with theoretical and numerical methods in random hypergraphs. We set up evolution equations for the pruning process, and self-consistency equations for the percolation properties. Based on our theory, we find that the pruning process generates a hybrid percolation transition for either $k\ge 3$ \textit{or} $q\ge 3$. The critical exponents obtained theoretically are confirmed with finite-size scaling analysis. Next, when $k=q=2$, we obtain a unconventional degree-dependent critical relaxation dynamics analytically and numerically. 
Finally, we apply the $(k,q)$-core decomposition to a real coauthorship dataset and recognize the leading groups at an early stage. 
\end{abstract}



\end{frontmatter}

\section{Introduction}

Recently, structural and dynamical properties of complex networks are being reconsidered in the framework of a hypergraph and simplicial complex to reflect individual-group relationships in a real situation. Here, the group consists of more than two elements, and the interaction among them is called higher-order interaction (HOI). Examples include the coauthorship of a research paper, the synapse of neurons in the human brain, and functional complexes in biological networks such as metabolic and protein interaction networks. Such a group interaction differs from a linear combination of pairwise interactions. For instance, if a protein is missing in a functional complex, then that functional complex does not work at all~\cite{Battiston_2020_HigherOrderReview,Battiston_2021_HigherOrder_NatPhys,Torres_2021_SIAMreview,Petri_2018_SimplicialActivityDrivenModel,Iacopini_2019_SimplicialSocialContagion,Jhun_2019_SISinHypergraph,Arenas_2020_PRR_EpidemicsInSimplicialComplex,Carletti_2020_PRE_RWonHypergraphs,Arenas_2019_PRL_OscillatorSimplexes,Skardal_2020_CommPhys_HigherOrderSync,Bianconi_2020_PRL_SimplicialKuramoto,Mulars_2020_PRE_MasterStabilityOnHypergraphs,Latora_2021_NHB_EvolutionOfSocialNetworks,Jongshin_2021_CHA_BC,Giusti_2016_Neuroscience_Simplex,Jost_2019_HypergraphChemicalReaction}.

A hypergraph consists of nodes (elements) and hyperedges (such as social groups, synapse, and functional complexes). The number of elements in a group is called hyperedge size and is denoted as $n_i$ with hyperedge index $i$. A node can relate to more than one hyperedge. Therefore, hyperedges are connected, as shown in Fig.~\ref{fig:pruing_illustration}. In contrast, a graph consists of nodes and links representing pairwise interactions.

Emerging phenomena created in the HOI representation can be different from those in the graph representation. For instance, oscillators governed by the Kuramoto model with HOIs can exhibit an explosive synchronization transition. However, oscillators with pair-wise interactions exhibit a continuous synchronization transition~\cite{Arenas_2019_PRL_OscillatorSimplexes,Skardal_2020_CommPhys_HigherOrderSync,Bianconi_2020_PRL_SimplicialKuramoto}.
The conventional epidemic models such as susceptible-infected-susceptible (SIS) and susceptible-infected-recovered (SIR) models exhibit continuous transitions when interactions are pairwise. However, they can exhibit discontinuous transitions when interactions are higher-order~\cite{Petri_2018_SimplicialActivityDrivenModel,Iacopini_2019_SimplicialSocialContagion,Jhun_2019_SISinHypergraph,Arenas_2020_PRR_EpidemicsInSimplicialComplex}.
Moreover, percolation transitions with HOI can be discontinuous~\cite{Bianconi_PRE2021_Higher-Order_Multiplex_Percolation,Coutinho_PRL2020_GRL}. Therefore, due to HOIs, various phase transitions in hypergraphs can have different types of transitions such as discontinuous and hybrid transitions.

$k$-core in the graph is the maximal subgraph in which all nodes have degrees of at least $k$~\cite{Newman_2018_Networks}.
The simplest way to obtain a $k$-core is recursively pruning all nodes with degrees less than $k$~\cite{Seidman_1983_k_core}. This pruning process decomposes a graph into a core subgraph in a hierarchical manner. This $k$-core decomposition has been applied to various issues~\cite{Kong_2019_PhysRep_kCore}. For example, it serves as an efficient way to identify influential spreaders in epidemic contagion~\cite{Stanley_NatPhys2010_influential_spreaders} and vulnerable (or robust) nodes in an electrical power grid system~\cite{Motter_Science2017_Vulnerable_set_in_Power_grid}. Additionally, it is applied in many areas such as protein interactions~\cite{Wuchty_2005_Protein}, evolution~\cite{KLIMEK_2009_Evolution}, neuron systems.~\cite{Schwarz_2006_EPL_Jamming}.

$k$-core percolation was studied in random graphs: Once an Erd\H{o}s-R\'enyi (ER) random network of size $N$ with mean degree $z$ is generated, all nodes with degrees less than $k$ are deleted recursively until no more nodes with degrees less than $k$ remain in the system. The fraction of nodes remaining in the largest $k$-core subgraph is defined as the order parameter $m$. The mean degree $z$ plays a role as the control parameter. The order parameter $m$ is of $O(1)$ for $z > z_c$, where $z_c$ is a transition point and decreases continuously with $z$. As $z$ approaches $z_c$, when $k \ge 3$, the deletion of a node from the ER network can lead to collapse of the giant $k$-core subgraph. Thus, the order parameter is expressed as
\begin{equation}\label{eq:order_parameter_hybrid_general}
m=
\begin{cases}
0 & {\rm for}~~ z < z_c \\
m_0 + a(z-z_c)^{\beta} & {\rm for}~~ z > z_c,
\end{cases}
\end{equation}
where $m_0$ and $a$ are constants, and $\beta=1/2$. The order parameter is discontinuous at the transition point and exhibits critical behavior. Thus, $k$-core percolation exhibits a hybrid phase transition~\cite{Dorogovtsev_2006_PRL_kCore,Goltsev_2006_PRE_kCore,Dorogovtsev_2006_PhysicaD_kcore,Baxter_2011_PRE_Bootstrap_vs_Kcore,Baxter_PRX_kCore,Kahng_2016_PRE_kCore}.

In this study, we extend the $k$-core decomposition in the graph into $(k,q)$-core decomposition in the hypergraph. With this, previous researchers considered the following: In Ref.~\cite{Bianconi_PRE2021_Higher-Order_Multiplex_Percolation}, $k$-core percolations for vertex and hyperedge are considered separately. This corresponds to the $(k,1)$- or $(1,q)$-core decomposition in our case below. Ref.~\cite{Coutinho_PRL2020_GRL} proposed another type of decomposition process by removing vertices with degree one and their neighbors repeatedly. However, this core cannot be considered a variant of the $(k,q)$-core we consider here.

In hypergraph, $(k,q)$-core decomposition can be proposed easily by converting a hypergraph into a bipartite network comprising nodes and hyperedges. In this bipartite representation, each hyperedge is connected to nodes that are the elements of the hyperedge. We call the connection between the pair of a node and a hyperedge \textit{bipartite link}. Intraedges between two nodes or two hyperedges are not permitted. $(k,q)$-core is the subgroup of the given hypergraph in which every node (hyperedge) has more than or equal to $k$ ($q$) degrees. $(k,q)$-core was initially proposed in computer science~\cite{Ahmed_2007_pqCore,Cerinsek_2015_ijCore,Liu_2020_ijCore} as a visualization tool for large-scale hypergraph data. However, those studies did not deal with phase transitions. $(k,q)$-core percolation is similarly defined to $k$-core percolation. We will consider the phase transition of $(k,q)$-core percolation.

A $(k,q)$-core substructure in the hypergraph is not similar to the corresponding $k$-core in the graph. We will show that the $(k,q)$-core decomposition is more efficient than $k$-core decomposition to identify an influential group such as leading scholar and super spreader groups. Moreover, the structure of the $(k,q)$-core enables us to understand what extent the information propagates or disease spreads across hyperedges~\cite{LaurentHD_2021_PRL_Social_Confinement,Aksoy2020}. This information is helpful for setting up a vaccination strategy.

This paper is organized as follows: In Sec.~\ref{sec:framework}, we first introduce and formulate $(k,q)$-core decomposition of random hypergraphs, in which node degrees and hyperedge sizes are not correlated and their distributions have finite moments. Additionally, we determine the transition type of $(k,q)$-core percolation analytically and numerically. It exhibits a hybrid phase transition. The critical behavior of the hybrid transition is calculated. In Sec.~\ref{sec:numerical}, the relaxation dynamics of the fraction of vertices with a given degree are considered at a transition point. The relaxation dynamics decays in a power-law manner. The associated exponent values are determined for degree and hyperedge size. In Sec.~\ref{sec:realworld}, we show that a redundant modular structure can be selected and eliminated using $(k,q)$-core decomposition in a real coauthorship hypergraph. In Sec.~\ref{discussion}, the summary and discussion are presented.

\section{\label{sec:framework}Theory of $(k,q)$-core decomposition and percolation}

\subsection{Definition and pruning process}
A hypergraph $G(V,E)$ comprises a pair of sets: set $V$ for nodes and set $E\in 2^V$ for hyperedges. A bipartite links connect pairs of vertices and hyperedges. Degree $z_i$ of vertices $v_i\in V$ means the number of hyperedges connected to vertex $v_i$, and size $n_i$ of hyperedge $e_i\in E$ refers to the number of vertices it contains.

\begin{figure*}[ht!]
    \includegraphics[width=\linewidth]{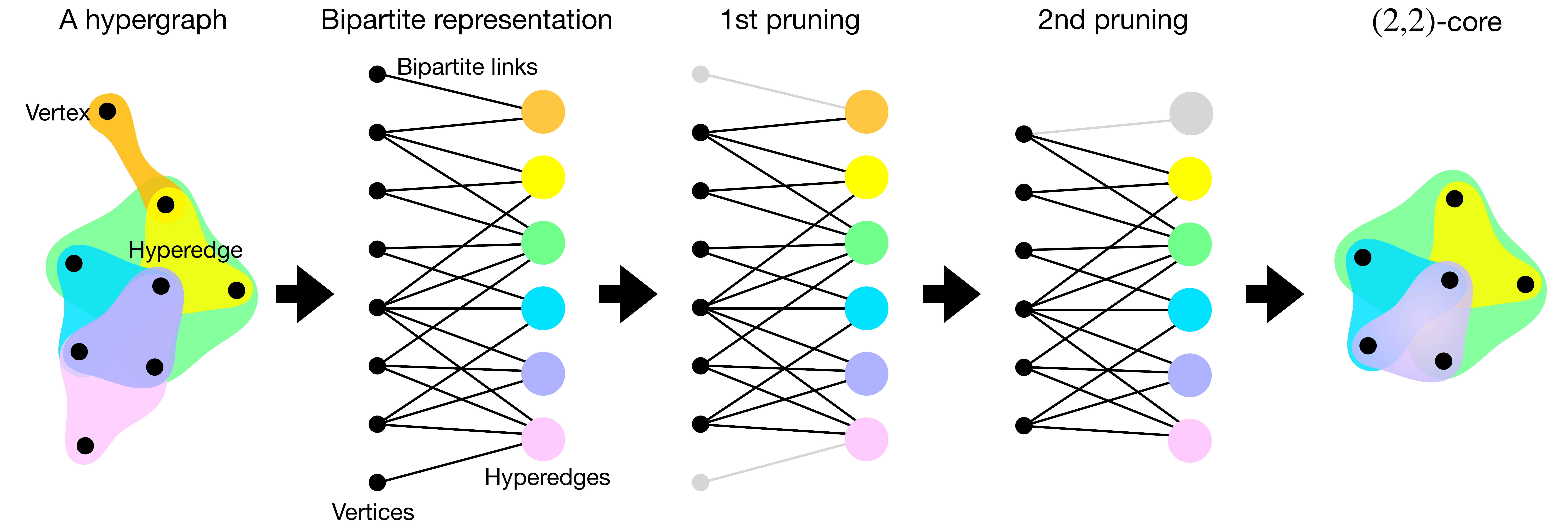}
    \caption{\label{fig:pruing_illustration} Illustration of $(2,2)$-core pruning process. Bipartite representation of a hypergraph is introduced for better understanding. In this representation, we simultaneously remove nodes (denoted as small circles) and hyperedges (denoted as large circles) of a degree and size less than two ($k=2$ and $q=2$) at each pruning step. A pruning process is repeated until no more nodes or hyperedges are smaller than two degrees and sizes. The remaining subgraph is the $(2,2)$ core of the given hypergraph in bipartite representation.}
\end{figure*}

We can obtain the $(k,q)$-core of a hypergraph by following the algorithm called $(k,q)$-core decomposition (see Fig.~\ref{fig:pruing_illustration}): For a hypergraph, all nodes with degrees less than $k$ and all hyperedges with size less than $q$ at each pruning step are removed. These removals may lead a few vertices and hyperedges to have remaining degrees less than $k$ and $q$, respectively. Additionally, these vertices and hyperedges are removed recursively until no more removal is required. The remaining nodes and hyperedges form the $(k,q)$-core of the hypergraph after a pruning.

In this work, we will focus on the cases of $k\ge 2$ and $q\ge 2$. The $(1,1)$-core decomposition implies that no pruning process occurs. Thus, the (1,1)-core is the graph itself. Furthermore, $(k,1)$-core pruning begins with removing vertices with a degree less than $k$ and hyperedges that have become an empty set. Then, the remaining vertices with at least degree $k$ or higher remain forever because they cannot make their hyperedges into an empty set. Similarly, $(1,q)$-core pruning first annihilates the hyperedges with a size smaller than $q$ and the vertices that do not belong to any hyperedge. Then, the remaining hyperedges of size $q$ or higher cannot be removed. Thus, the pruning process for either $k=1$ or $q=1$ can proceed at most up to one cycle (two steps) even in the case of infinite-sized hypergraphs.

$(k,2)$-core decomposition is equivalent to the traditional $k$-core decomposition of graphs. In the $k$-core pruning process, the edge connected to a vertex is removed when the vertex is pruned so that self-loops cannot exist. In contrast, for the $(k,1)$-core decomposition, removing vertices from the hypergraph may result in edges containing only one vertex, which can be considered self-loops. Therefore, the $(k,2)$-core decomposition, which excludes edges with only one vertex, describes the $k$-core decomposition of graphs.

\subsection{Evolution equations for uncorrelated hypergraphs}
Here, we construct the evolution equations of the degree and size distributions in the pruning process introduced in Ref.~\cite{Baxter_PRX_kCore}. A hypergraph is represented by a bipartite network comprising vertices and hyperedges. First, consider $P^{(v)}(z,t)$ to be the fraction of vertices having degree $z$ at time $t$ and $P^{(e)}(n,t)$ be the fraction of hyperedges whose size is $n$ at time $t$. Here, $t$ is the time step of the pruning process. At each time step, all nodes with degrees less than $k$ and hyperedges with a size less than $q$, which are called disqualified, are pruned by removing all bipartite links connected to them.
Furthermore, consider $R_k^{(v)}(t)$ $(R_q^{(e)}(t))$ to be the probability of reaching a disqualified vertex (hyperedge) at time $t$ at the end of a bipartite link randomly selected. They are evaluated in terms of $P^{(v)}(z,t)$ and $P^{(e)}(n,t)$ as follows:

\begin{eqnarray}\label{eq:exact_removal_prob}
\begin{aligned}
R_k^{(v)}(t)=& \frac{1}{\langle z(t) \rangle^{(v)}}\sum_{z<k} z P^{(v)}(z,t) \,,\\
R_q^{(e)}(t)=& \frac{1}{\langle n(t) \rangle^{(e)}}\sum_{n<q} n P^{(e)}(n,t) \,.\\
\end{aligned}
\end{eqnarray}
with $\langle z(t)\rangle^{(v)}$ and $\langle n(t)\rangle^{(e)}$ denoting the mean degree and mean hyperedge size, respectively, at time step $t$.
These formulae were derived under the assumption that no correlation exists between node degrees and hyperedge sizes.
The probability that a node of degree $ z^\prime \ge k$ at time $t$ has degree $z$ at time $t+1$ is written as
\begin{eqnarray}
    T_{z'z}^{(v)}(t)\equiv\binom{z'}{z} (1 - R_q^{(e)}(t))^z R_q^{(e)}(t)^{z' -z}    \,.
\end{eqnarray}
Similarly, the probability that a hyperedge of size $ n'\ge q$ at time $t$ has hyperedge size $n$ at $t+1$ is
\begin{eqnarray}
    T_{n'n}^{(e)}(t)\equiv\binom{n'}{n} (1 - R_k^{(v)}(t))^n R_k^{(v)}(t)^{n' -n} \,.
\end{eqnarray}
Summing over all possible degrees $z'$ and sizes $n'$, we obtain the degree and size distributions evolved by the following evolution equations:
\begin{eqnarray}
\begin{aligned}
P^{(v)}&(z,t+1) = \sum_{z'\ge \max\lbrace z,k\rbrace} P^{(v)}(z',t) T_{z'z}^{(v)}(t)\,,\\
P^{(e)}&(n,t+1) = \sum_{n'\ge \max\lbrace n,q\rbrace} P^{(e)}(n',t) T_{n'n}^{(e)}(t)\,,
\end{aligned}\label{eq:exact_evolution_equations}
\end{eqnarray}
for $z>0$ and $n>0$. In contrast, when $z=0$ and $n=0$, the equations become
\begin{eqnarray}
\begin{aligned}
P^{(v)}(0,t+1) =& \sum_{z'<k} P^{(v)}(z',t) +  \sum_{z'\ge k} P^{(v)}(z',t) T_{z'0}^{(v)}(t)\,,\\
P^{(e)}(0,t+1) =& \sum_{n'<q} P^{(e)}(n',t) + \sum_{n'\ge q} P^{(e)}(n',t) T_{n'0}^{(e)}(t)\,.
\end{aligned}\label{eq:exact_evolution_equations_for_zero}
\end{eqnarray}
Recall that the probabilities $P^{(v)}(0,t)$ and $P^{(e)}(0,t)$ represent the accumulated fractions of nodes and hyperedges removed during the pruning process up to time $t$ and those with all neighbors pruned at time $t$.

We remark on a few things. First, as mentioned in the previous section, $P^{(e)}(n) = \delta_{n,2}$ and $q = 2$ for graph pruning. Second, the previous statement that the pruning process for the $(k,1)$-core (the $(1,q)$-core) can proceed up to one step is reflected in the fact that $R_1^{(v,e)}=0$ from Eq.~\eqref{eq:exact_removal_prob}.

\subsection{$(k,q)$-core percolation}\label{sec:kq_percolation_theory}

The number of vertices surviving the pruning process may or may not be as large as that of $O(N)$, indicating whether the $(k,q)$-core percolation occurs. As in Ref.~\cite{Dorogovtsev_2006_PRL_kCore}, we consider that a giant $(k,q)$-core, the core including a number $O(N)$ of vertices, is the part of tree structure around the critical point and present an analytical approach to explore the $(k,q)$-core percolation. It will be applied to specific model hypergraphs in the next section.

\subsubsection{Self-consistent equations}
A vertex (hyperedge) belongs to a giant $(k,q)$-core if it is connected by at least $k \, (q)$ links to certain hyperedges (vertices) of the giant core. Introducing $R$ and $H$ to denote the probability that the end vertex and hyperedge, respectively, of a randomly selected bipartite link does not belong to a giant core, we can set up the equations for the probability $m^{(v)} (m^{(e)})$ that a vertex (hyperedge) belongs to a giant $(k,q)$-core as
\begin{align}
m^{(v)} &= \sum_{r=k}^{\infty} \sum_{z=r}^\infty P^{(v)}(z) \binom{z}{r}(1-H)^r H^{z-r},\nonumber\\
m^{(e)} &= \sum_{s=q}^{\infty} \sum_{n=s}^\infty P^{(e)}(n) \binom{n}{s}(1-R)^s R^{n-s}.
\label{eq:m}
\end{align}
$P^{(v)}(z)$ and $P^{(e)}(n)$ are the original degree and size distribution of a given hypergraph before pruning, i.e., $P^{(v)}(z) = P^{(v)}(z,t=0)$ and $P^{(e)}(n) = P^{(n)}(n,t=0)$, respectively, with the notations in Sec. II B used.
Similarly, considering that a link cannot lead to a giant core via its end vertex (hyperedge) if the end vertex (hyperedge) has at most $k-2 \, (q-2)$ additional links leading to the giant core, we obtain that $R$ and $H$ satisfy
\begin{equation}
R = \Phi_k^{(v)}(H) \  {\rm and} \ H = \Phi_q^{(e)}(R),
\end{equation}
and thus, can be obtained by solving
\begin{align}
R &= \Theta_{k,q}^{(v,e)}(R) \equiv \Phi_k^{(v)} \left(\Phi_q^{(e)}(R)\right),\nonumber\\
H &= \Theta_{q,k}^{(e,v)} (H)\equiv \Phi_q^{(e)} \left(\Phi_k^{(v)}(H)\right),
\label{eq:RH}
\end{align}
where
\begin{align}
\Phi_k^{(v)}(x) &\equiv \sum_{r=0}^{k-2} \sum_{z=r}^\infty {(z+1) P^{(v)}(z+1)\over \langle z\rangle^{(v)}} \binom{z}{r} (1-x)^r x^{z-r}\nonumber\\
&= \sum_{r=0}^{k-2} {(1-x)^r \over r!} {\text{d}^r\over \text{d}x^r} G_1^{(v)}(x), \nonumber\\
\Phi_q^{(e)}(x) &\equiv \sum_{s=0}^{q-2} \sum_{n=s}^\infty {(n+1) P^{(e)}(n+1)\over \langle n\rangle^{(e)}} \binom{n}{s} (1-x)^s x^{n-s}\nonumber\\
&= \sum_{s=0}^{q-2} {(1-x)^s \over s!} {\text{d}^s\over \text{d}x^s} G_1^{(e)}(x)
\label{eq:Phi}
\end{align}
with $G_1^{(v)}(x)\equiv \sum_{z=0}^\infty z P^{(v)}(z) x^{z-1}/\langle z\rangle^{(v)}$ and $G_1^{(e)}(x)\equiv \sum_{n=0}^\infty n P^{(e)}(n) x^{n-1}/\langle n\rangle^{(e)}$ being the generating functions of the degree and size distributions, respectively.

\subsubsection{Nature of the $(k,q)$-core percolation}
Solving first Eq.~(\ref{eq:RH}) for $R$ and $H$ and using the solution in Eq.~(\ref{eq:m}), we can obtain the probability of a vertex and a hyperedge belonging to the giant $(k,q)$-core, the order parameter of the $(k,q)$-core percolation. It is obvious that $R=H=1$ is the solution to Eq.~(\ref{eq:RH}). If it is the only solution, then $m^{(v)} =0$ and $m^{(e)}=0$ by Eq.~(\ref{eq:m}); i.e., there is no giant $(k,q)$-core. If a non-trivial solution $R<1$ and $H<1$ exists, it will give a positive value of $m^{(v)}$ and $m^{(e)}$. Examining the behaviors of $\Phi_k^{(v)}(x)$ and $\Phi_q^{(e)}(x)$ around $x=1$ that depend on $k$ and $q$, we can understand the non-trivial solutions and the nature of the $(k,q)$-core percolation.

Expanding Eq.~(\ref{eq:Phi}) around $x=1$ and using the relation $\sum_{r=0}^b \binom{a}{r} (-1)^r = (-1)^b \binom{a-1}{b}$, we obtain 
\begin{align}
\Phi_k^{(v)}(x) &= 1  - {1\over \Gamma(k-1)} \sum_{r = k-1}^\infty {(-1)^{r-k+1} \over \Gamma(r-k+2)} {\langle z_{r+1}\rangle^{(v)}\over \langle z\rangle^{(v)}} {(1-x)^r \over r}\nonumber\\
\Phi_q^{(e)}(x) &= 1  - {1\over \Gamma(q-1)} \sum_{s = q-1}^\infty {(-1)^{s-q+1} \over \Gamma(s-q+2)} {\langle n_{s+1}\rangle^{(e)}\over \langle n\rangle^{(e)}} {(1-x)^s \over s}
\label{eq:PhiExpand}
\end{align}
with $\Gamma(x)$ being the Gamma function and the factorial moments of the degree distribution $\langle z_r\rangle^{(v)} \equiv \sum_{z=0}^\infty z (z-1) (z-2) \cdots (z-r+1) P^{(v)}(z)$ and of the size distribution $\langle n_s\rangle^{(e)} \equiv \sum_{n=0}^\infty n (n-1) (n-2) \cdots (n-s+1) P^{(e)}(n)$ used. We consider that all the factorial moments of the degree and size distributions are finite. The leading-order behaviors of $\Theta_{k,q}^{(v,e)}(x)$ and $\Theta_{q,k}^{(e,v)}(x)$ in Eq.~(\ref{eq:RH}) near $x=1$ are given by \begin{align}
\Theta_{k,q}^{(v,e)} (x) & \simeq 1 - {\langle z_k\rangle^{(v)} \over \Gamma(k) \langle z\rangle^{(v)}}\left({\langle n_q\rangle^{(e)} \over \Gamma(q) \langle n\rangle^{(e)}}\right)^{k-1} (1-x)^{(q-1)(k-1)}\nonumber\\
&+O((1-x)^{q(k-1)}, (1-x)^{k(q-1))}),
\nonumber\\
\Theta_{q,k}^{(e,v)} (x) & \simeq 1 - {\langle n_q\rangle^{(e)} \over \Gamma(q) \langle n\rangle^{(e)}}\left({\langle z_k\rangle^{(v)} \over \Gamma(k) \langle z\rangle^{(v)}}\right)^{q-1} (1-x) ^{(q-1)(k-1)}\nonumber\\
&+O((1-x)^{q(k-1)}, (1-x)^{k(q-1))}).
\label{eq:PhiExpand2}
\end{align}

If $k=2$ and $q=2$, both $\Theta(x)$'s in Eq.~(\ref{eq:PhiExpand2}) decrease linearly near $x=1$ with $x$, and then reach $\Theta(0)>0$ at $x=0$. If the common coefficient of the linear terms in Eq.~(\ref{eq:PhiExpand2}) with $k=q=2$ is larger than one, i.e.,
\begin{equation}
{\langle z_2\rangle^{(v)} \over \langle z\rangle^{(v)}}
{\langle n_2\rangle^{(e)} \over \langle n\rangle^{(e)}} =
{\langle z(z-1)\rangle^{(v)} \over \langle z\rangle^{(v)}}
{\langle n(n-1)\rangle^{(e)} \over \langle n\rangle^{(e)}} > 1,
\label{eq:critical22}
\end{equation}
then we can observe that $\Theta_{k,q}^{(v,e)}(x)<x$ is in the range $R<x<1$ and $\Theta_{q,k}^{(e,v)}(x)<x$ is in the range $H<x<1$, and these $R$ and $H$
 are the nontrivial solutions to Eq.~(\ref{eq:RH}). In contrast, if Eq.~(\ref{eq:critical22}) does not hold, $\Theta(x)$'s are larger than $x$ in the entire range $0<x<1$. From Eq.~(\ref{eq:critical22}), the critical manifold of the $(2,2)$-core percolation is given by $\dfrac{\langle z_2\rangle^{(v)} }{\langle z\rangle^{(v)}}\dfrac{\langle n_2\rangle^{(e)} }{\langle n\rangle^{(e)}} = 1$. Introducing
\begin{eqnarray*}
    \Delta \equiv {\langle z(z-1)\rangle^{(v)} \over \langle z\rangle^{(v)}}{\langle n(n-1)\rangle^{(e)} \over \langle n\rangle^{(e)}}-1\,,
\end{eqnarray*}
we obtain that when $\Delta$ is positive and small, $\Theta(x)$ in Eq.~(\ref{eq:PhiExpand2}) behaves near $x=1$ as
\begin{equation}
\Theta(x) -x  \sim - \Delta (1-x) + b (1-x)^2
\end{equation}
with $b=\dfrac{1}{2}\left\{ \dfrac{\langle z_2\rangle^{(v)}}{\langle z\rangle^{(v)}}\dfrac{\langle n_3\rangle^{(e)}}{\langle n\rangle^{(e)}}+\dfrac{\langle z_3\rangle^{(v)}}{\langle z\rangle^{(v)}}\dfrac{{\langle n_2\rangle^{(e)}}^2}{{\langle n\rangle^{(e)}}^2}\right\}$. Thus, in case of $\Theta_{2,2}^{(v,e)}(x)$, the solution of $\Theta(x)-x=0$ will be given by $x\simeq 1-{\Delta / b}$. This implies that the nontrivial solution to Eq.~(\ref{eq:RH}) behaves as
$1-R \sim \Delta$ and $1-H\sim \Delta$ for $0<\Delta\ll 1$. One can immediately see from Eq.~(\ref{eq:m}) that the order parameters are $m^{(v)}\sim (1-H)^2 \sim \Delta^2$ and $m^{(e)}\sim (1-R)^2 \sim \Delta^2$ for $0<\Delta\ll 1$. Therefore, we obtain for $k=q=2$ that
\begin{equation}
m^{(v)}\sim
\begin{cases}
0 & (\Delta<0) \\
\Delta^\beta & (0<\Delta\ll 1)
\end{cases}
\ {\rm with} \ \beta=2
\end{equation}
and the same holds for $m^{(e)}$. The results show that the $(2,2)$-core percolation is a continuous transition.

When either $k$ or $q$ is larger than $2$, the leading order $(q-1)(k-1)$ in Eq.~(\ref{eq:PhiExpand2}) is larger than one. Moreover, the derivative of $\Theta(x)$'s at $x=1$ is zero,
\begin{equation}
{\text{d}\Theta_{k,q}^{(v,e)}(x)\over \text{d}x}\bigg|_{x=1}=0 \quad {\rm and} \quad
{\text{d}\Theta_{q,k}^{(e,v)}(x)\over \text{d}x}\bigg|_{x=1}=0.
\end{equation}
This helps us grasp the behavior of $\Theta(x)$'s as follows. As $x$ decreases, $y=\Theta(x)$ decreases from $1$ quite slowly near $x=1$ and then decreases relatively fast to reach $\Theta(0)$ at $x=0$. Therefore, if a condition similar to Eq.~(\ref{eq:critical22}) is met, $\Theta_{k,q}^{(v,e)}(x)<x$ can be in the range $R<x<\tilde{R}$, $\Theta_{q,k}^{(e,v)}(x)<x$ can be in the range $H<x<\tilde{H}$, and $R$ and $H$ are the nontrivial solutions to Eq.~(\ref{eq:RH}). In contrast, $\tilde{R}$ and $\tilde{H}$, constants smaller than one, are unphysical solutions. If the condition is not met, $\Theta(x)>x$ in the entire range $0<x<1$.
 Consider that $\Delta$ can be defined similarly to the case of $k=q=2$ as a measure of the distance to the critical manifold. Additionally, $y=x$ is tangential to $y=\Theta(x)$ at $x_0$ when $\Delta=0$ with $x_0$ possibly not close to $1$. Then, when $\Delta$ is positive and small, $\Theta(x)$ is expected to behave near $x=x_0$ as
\begin{equation}
\Theta(x) - x \sim - \Delta +  c (x-x_0)^2
\end{equation}
with $c=\dfrac{1}{2}\dfrac{\text{d}^2 \Theta_{k,q}^{(v,e)}(x)}{\text{d}x^2}\bigg|_{x=x_0}$ or $c=\dfrac{1}{2}\dfrac{\text{d}^2 \Theta_{q,k}^{(e,v)}(x) }{\text{d}x^2}\bigg|_{x=x_0}$ and the solution to $\Theta(x)-x=0$ will be given by $x = x_0 - \left({\Delta}/{c}\right)^{1/2}$. This implies for $0<\Delta\ll 1$ that $R \simeq R_0 - {\rm (const.)} \Delta^{1/2}$ and $H \simeq H_0 - {\rm (const.)} \Delta^{1/2}$, and the order parameters are given by $m^{(v)}\sim m^{(v)}_0 + ({\rm const.}) \Delta^{1/2}$ and $m^{(e)}\sim m^{(e)}_0 + ({\rm const.}) \Delta^{1/2}$, where $m^{(v)}_0$ and $m^{(e)}_0$ are obtained from Eq.~(\ref{eq:m}) with $H_0$ and $R_0$ used. The order parameters for $k\geq 3$ or $q\geq 3$ are therefore given by
\begin{equation}\label{eq:order_parameter_hybrid}
m^{(v)}\sim
\begin{cases}
0 & (\Delta<0) \\
m^{(v)}_0 + ({\rm const.}) \Delta^{1/2} & (0<\Delta\ll 1)
\end{cases}
\end{equation}
and the same holds for $m^{(e)}$. The results show that the $(k,q)$-core percolation is a hybrid transition when $k\ge 3$ or $q\ge 3$~\cite{Dorogovtsev_2006_PRL_kCore,Goltsev_2006_PRE_kCore}.

\section{$(k,q)$-core of random hypergraphs}\label{sec:numerical}

We investigate the properties of $(k,q)$-core percolation on a random synthetic hypergraph of different degree distributions. We focus on the properties of vertices. However, the properties of hyperedges can be obtained similarly.

Random hypergraphs can be constructed by generalizing the rule of Erd{\H{o}}s--R\'enyi (ER) random graphs~\cite{ER_1960_evolution}.
Given system size $N$, the mean degree $\langle z\rangle^{(v)}$, and the mean hyperedge size $\langle n\rangle^{(e)}$, we consider $N$ vertices and $E = N \langle z\rangle^{(v)} / \langle n\rangle^{(e)}$ empty hyperedges first. Then, we select a pair of vertices and hyperedges randomly and add a bipartite link between the pair if they are not connected yet. This process is repeated until the number of bipartite links reaches $L=N\langle z\rangle^{(v)}=E\langle n\rangle^{(e)}$. Multiple bipartite links are not allowed, i.e., a hyperedge cannot contain the same vertex more than once.

This model generates Poisson distributions of both degrees $P^{(v)}(z) = {(\langle z\rangle^{(v)})^z}e^{-\langle z\rangle^{(v)}}/{z!}$ and hyperedge size $P^{(e)}(n) = {(\langle n\rangle^{(e)})^n}e^{-\langle n\rangle^{(e)}}/{n!}$ because each vertex (hyperedge) is selected independently with uniform probability $1/N$ ($1/E$).

\begin{figure*}[t!]
\includegraphics[width=\linewidth]{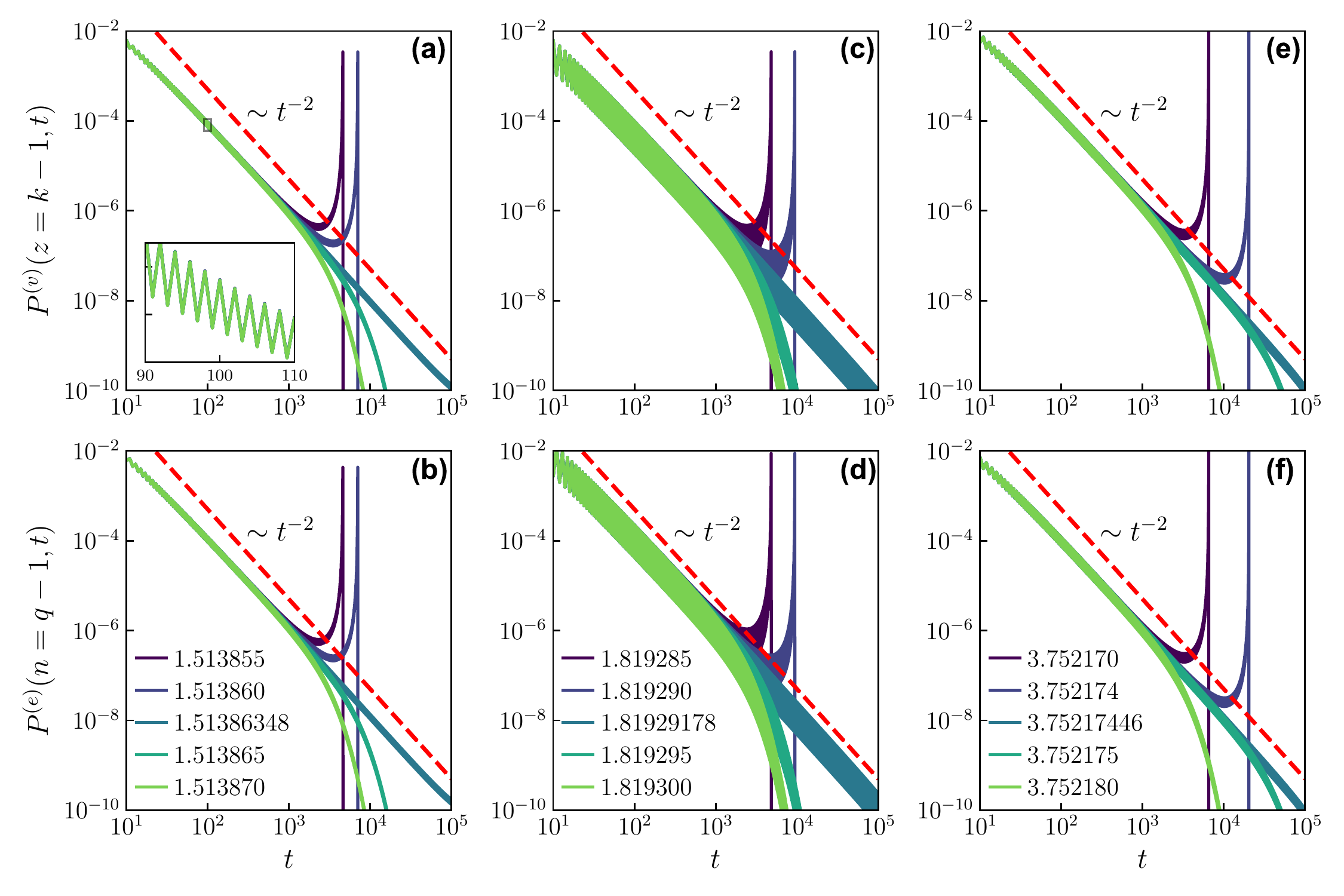}
\caption{\label{fig:power_law_decay} Plots of the distributions $P^{(v)}(k-1,t)$ and $P^{(e)}(q-1,t)$ vs time step $t$ at the transition point for various $k$ and $q$ pairs and mean node degrees $\langle z \rangle$. All data points are obtained from numerical solutions of the exact time evolution equations Eqs.~\eqref{eq:exact_evolution_equations} and \eqref{eq:exact_evolution_equations_for_zero}. (a) and (b) are for $(2,3)$-core pruning, (c) and (d) are for $(3,2)$-core pruning, and (e) and (f) are for $(3,3)$-core pruning processes. Legends denote the mean vertex degree $\langle z\rangle^{(v)}$. The mean hyperedge size is fixed as $\langle n\rangle^{(e)} = 3.0$. The red dashed guidelines represent $t^{-2}$ in all the plots. The inset of (a) shows a zoomed image displaying oscillating behavior. The $y$-axis is drawn on logarithmic scale.}
\end{figure*}

\subsection{Relaxation of degree and size distributions}
We investigate the relaxation dynamics of vertex degrees and hyperedge sizes during the pruning process at a transition point. As the pruning process proceeds, the structure of the network changes. Thus, the fraction of vertices (hyperedges) with degree $z$ (size $n$) at time $t$, denoted as $P^{(v)}(z,t)$ ($P^{(e)}(n,t)$), may change with time by Eqs.~\eqref{eq:exact_evolution_equations} and \eqref{eq:exact_evolution_equations_for_zero}. Similar to the dynamics in the $k$-core decomposition~\cite{Baxter_PRX_kCore}, the primary disqualified vertices of degree $k-1$ and hyperedges of size $q-1$ drive the pruning dynamics. Additionally, we numerically obtain that $P^{(v)}(z=k-1,t)$ decays with time in a power-law manner as $P^{(v)}(k-1,t)\sim t^{-2}$ at the transition point $\langle z\rangle_c^{(v)}$ for $k \ge 3$ and given $\langle n\rangle^{(e)}$. $P^{(e)}(n=q-1,t)$ exhibits a power-law behavior as $P^{(e)}(q-1,t)\sim t^{-2}$ at the transition point $\langle n\rangle_c^{(e)}$ for $q \ge 3$, given $\langle z\rangle^{(v)}$, as shown in Fig.~\ref{fig:power_law_decay}.

\begin{figure*}[h!tb]
\includegraphics[width=\linewidth]{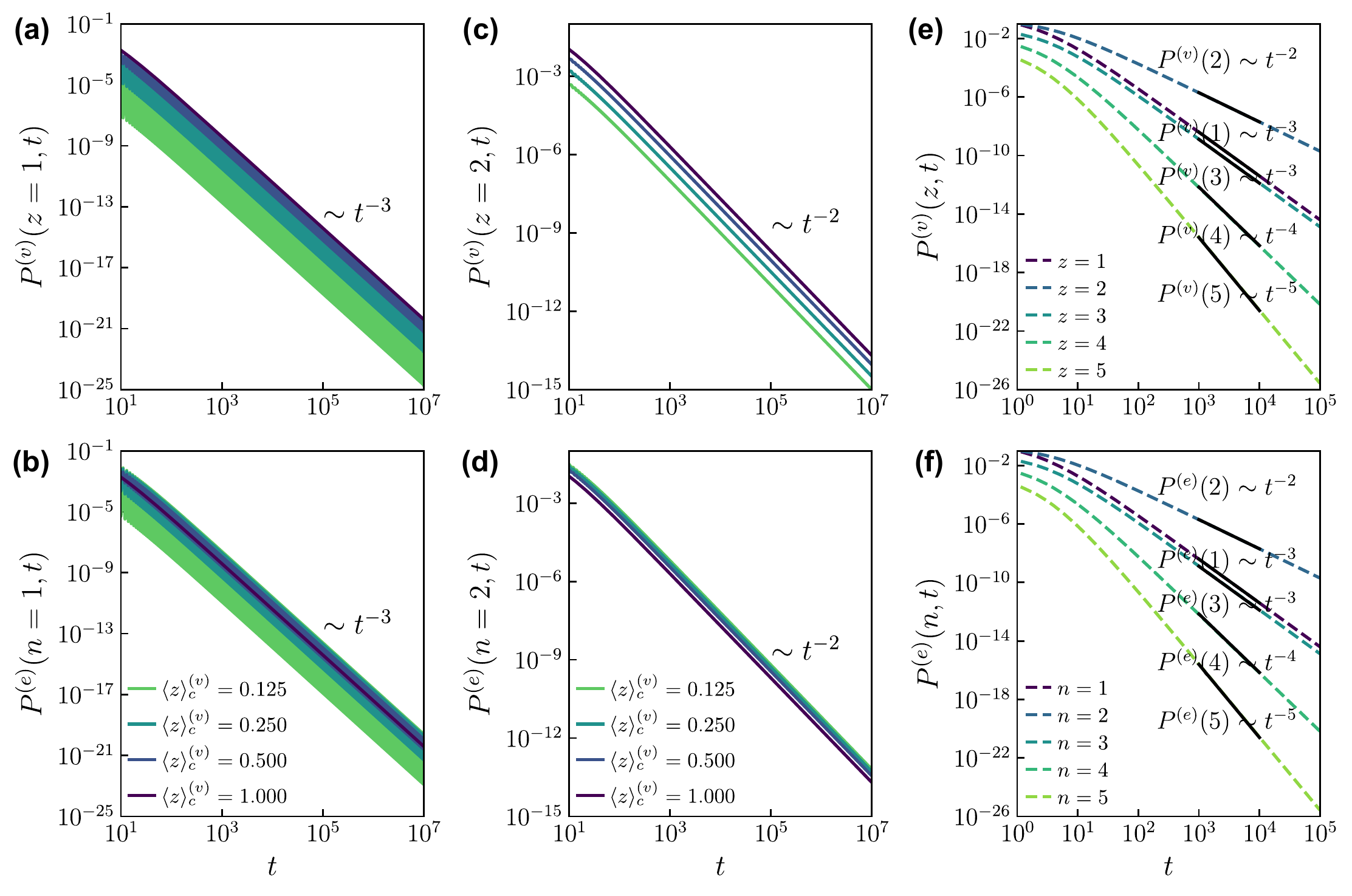}
\caption{\label{fig:2core_power_law_degree_evolution} Plots of the distributions of the fraction of vertices with degree $z$, denoted as $P^{(v)}(z,t)$ and the fractions of hyperedges of size $n$ denoted as $P^{(e)}(n,t)$ in $(2,2)$-core decomposition for random hypergraphs at the transition points for the vertices of degree 1 (a) and 2 (c), and for the hyperedge of size 1 (b) and 2 (d). (a)-(d) $\langle z \rangle^{(v)}$ and $\langle n \rangle^{(e)}$ are taken on the critical manifold $\langle z \rangle_c^{(v)} \langle n \rangle_c^{(e)}=1$, and $\langle z \rangle_c^{(v)}$ is given in legend in (b) and (d). (e) Plot of $P^{(v)}(z,t)$ vs $t$ for general $z$. (f) Plot of $P^{(e)}(n,t)$ vs $t$ for general $n$. We consider $\langle z \rangle^{(v)} = \langle n \rangle^{(e)} =1$}
\end{figure*}

Different from the case of $k\geq 3$ or $q\geq 3$, the decay of the fractions of the primary disqualified vertices and hyperedges at the transition point for $k=q=2$ is characterized by a power-law exponent $-3$, not $-2$, as
\begin{equation}
P^{(v)}(z=k-1=1,t)~\sim t^{-3} \ {\rm and} \ P^{(e)}(n=q-1=1,t)\sim t^{-3},
\label{eq:P1_22}
\end{equation}
as shown in Fig.~\ref{fig:2core_power_law_degree_evolution}. As we have shown in Sec.~\ref{sec:framework} and will present numerical results in the next subsection, the $(2,2)$-core percolation transition is continuous, while it is a hybrid type for $k\geq 3$ or $q\geq 3$. The $(2,2)$-core is not as large as $O(N)$ at the transition point. Therefore, the fractions of the vertices of any nonzero degree and the hyperedges of any nonzero size vanish in the long-time limit. In contrast, $P^{(v)}(z,t)$ and $P^{(e)}(n,t)$ are nonzero for $z\geq k$ and $n\geq q$ if $k\geq 3$ or $q\geq 3$~\cite{Baxter_PRX_kCore}.
Considering this aspect and that pruning propagates in a tree structure as in Ref.~\cite{Baxter_PRX_kCore}, we can obtain the following by solving Eqs.~\eqref{eq:exact_evolution_equations} and ~\eqref{eq:exact_evolution_equations_for_zero} 
\begin{equation}
P^{(v)}(z\geq 2,t) \sim t^{-z} \ {\rm and} \ P^{(e)}(n\geq 2,t) \sim t^{-n}
\label{eq:Pgeneral_22}
\end{equation}
at the transition point. We present the derivation of Eqs.~(\ref{eq:P1_22}) and (\ref{eq:Pgeneral_22}) in~\ref{Appendix:Analytic_calculation_for_22core}. Such different scaling behaviors depending on $z$ and $n$ are the primary novel results of the present work.

The transition point can be determined as the point where $P^{(v)}(k-1,t)$ and $P^{(e)}(q-1,t)$ exhibit power-law decay behavior. In Fig.~\ref{fig:power_law_decay}, for $\langle n\rangle^{(e)}=3.0$, we estimate the critical point: for the $(2,3)$ core, $\langle z\rangle_c^{(v)} = 1.513\, 863\, 48$; for the $(3,2)$ core, $\langle z\rangle_c^{(v)} = 1.819\, 291\, 78$; and for the $(3,3)$ core, $\langle z\rangle_c^{(v)} = 3.752\, 174\, 46$. These precisely estimated values of $\langle z\rangle_c^{(v)}$ help in estimating the critical exponents using finite-size scaling analysis in the next subsection.

\subsection{$(k,q)$-core percolation}
Using the Poisson degree and size distributions, we can obtain the self-consistency functions in Eq.~\eqref{eq:Phi} for random hypergraphs as
\begin{align}
        \Phi_k^{(v)}(x) =& \sum_{r=0}^{k-2}\frac{(1-x)^r}{r!\langle z\rangle^{(v)}} \frac{\text{d}^{r+1}}{\text{d}x^{r+1}}e^{-\langle z\rangle^{(v)}(1-x)}\nonumber\\
        =&e^{-(1-x)\langle z\rangle^{(v)}}\sum_{r=0}^{k-2}\frac{\lbrace\langle z\rangle^{(v)}(1-x)\rbrace^r}{r!}\nonumber\\
        =& \frac{\Gamma(k-1,\langle z\rangle^{(v)}(1-x))}{\Gamma(k-1)}\,,
\end{align}
and
\begin{eqnarray}
        \Phi_q^{(e)}(x) =\frac{\Gamma(q-1,\langle n\rangle^{(e)}(1-x))}{\Gamma(q-1)}\,,
\end{eqnarray}
which leads to
\begin{align}
\Theta_{k,q}(x) =& \Phi_k^{(v)}\left(\Phi_q^{(e)}(x)\right) \nonumber\\
=& \frac{1}{\Gamma(k-1)}\Gamma\left(k-1,\frac{\Gamma(q-1,(1-x)\langle n\rangle^{(e)})\langle z\rangle^{(v)}}{\Gamma(q-1)}\right)\,,
\end{align}
where $\Gamma(n,x)=\int_{x}^\infty \text{d}t t^{n-1}e^{-t}$ is the incomplete Gamma function.
The nature of the $(k,q)$-core percolation transition can be understood by analyzing the behavior of $\Theta_{k,q}(x)$ as presented in Sec.~\ref{sec:kq_percolation_theory}. We present the behavior of the self-consistency function $\Theta_{k,q}^{(v,e)}(x)$ in~\ref{Appendix:Self_Consistency}.

\begin{figure*}[h!tb]
\includegraphics[width=\linewidth]{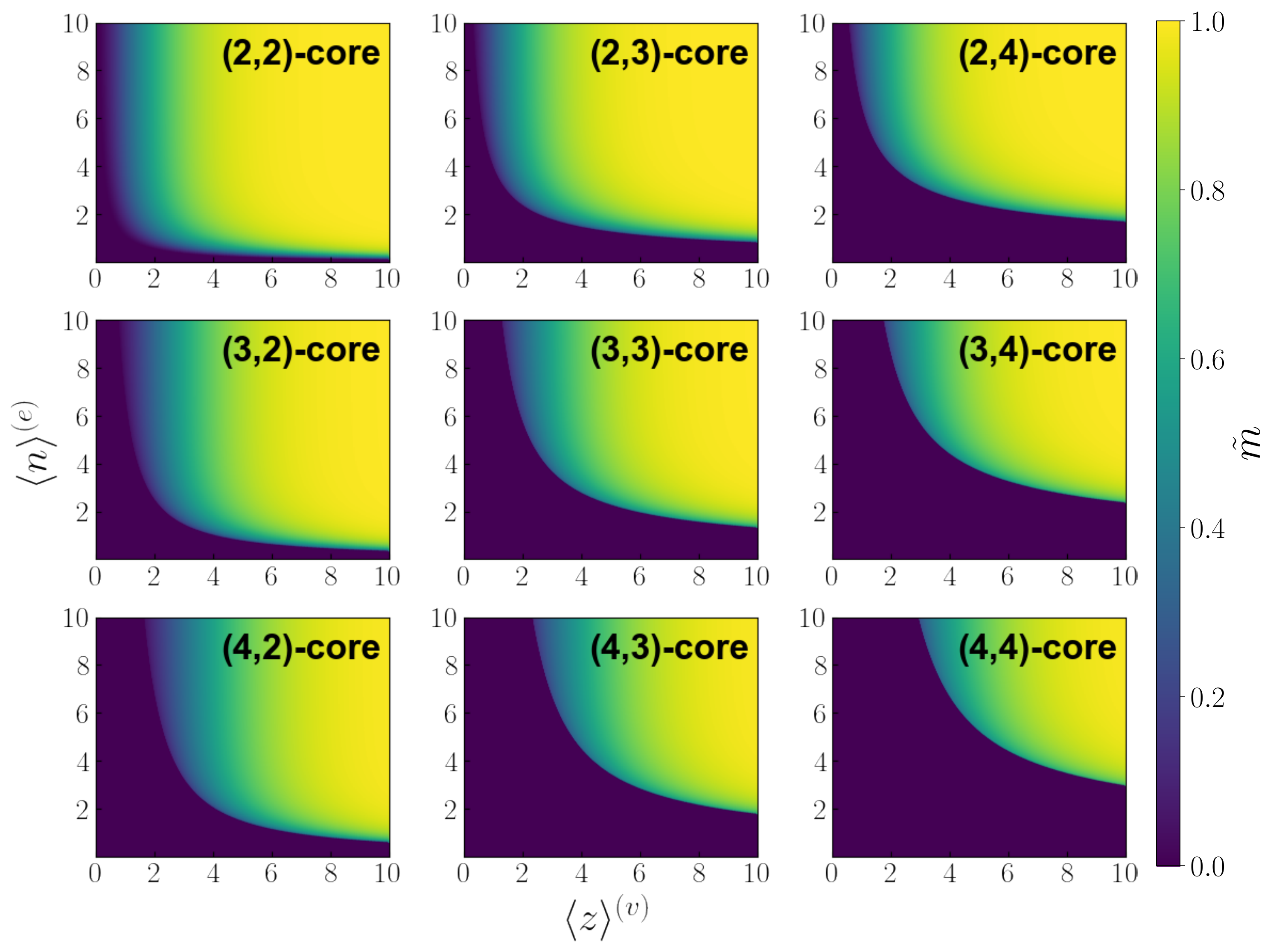}
\caption{\label{fig:phase_diagram_theory} Phase diagrams of the percolation transition of the random hypergraphs for various $(k,q)$ pairs. The system size we used is $N=10^5$ and the number of ensembles for each point is $10^3$. Here, the fraction $\tilde{m}$ of vertices that remain throughout the $(k,q)$-core decomposition for a given mean node degree $\langle z \rangle^{(v)}$ and mean hyperedge size $\langle n \rangle^{(e)}$ are plotted. For each point in the diagram, the distributions of degree $P^{(v)}(z) $ and hyperedge sizes $P^{(e)}(n)$ follow the Poisson distributions with the averages of the corresponding control parameter $\langle z\rangle^{(v)}$ and $\langle n\rangle^{(e)}$.}
\end{figure*}

\begin{figure*}[h!tb]
\includegraphics[width=\linewidth]{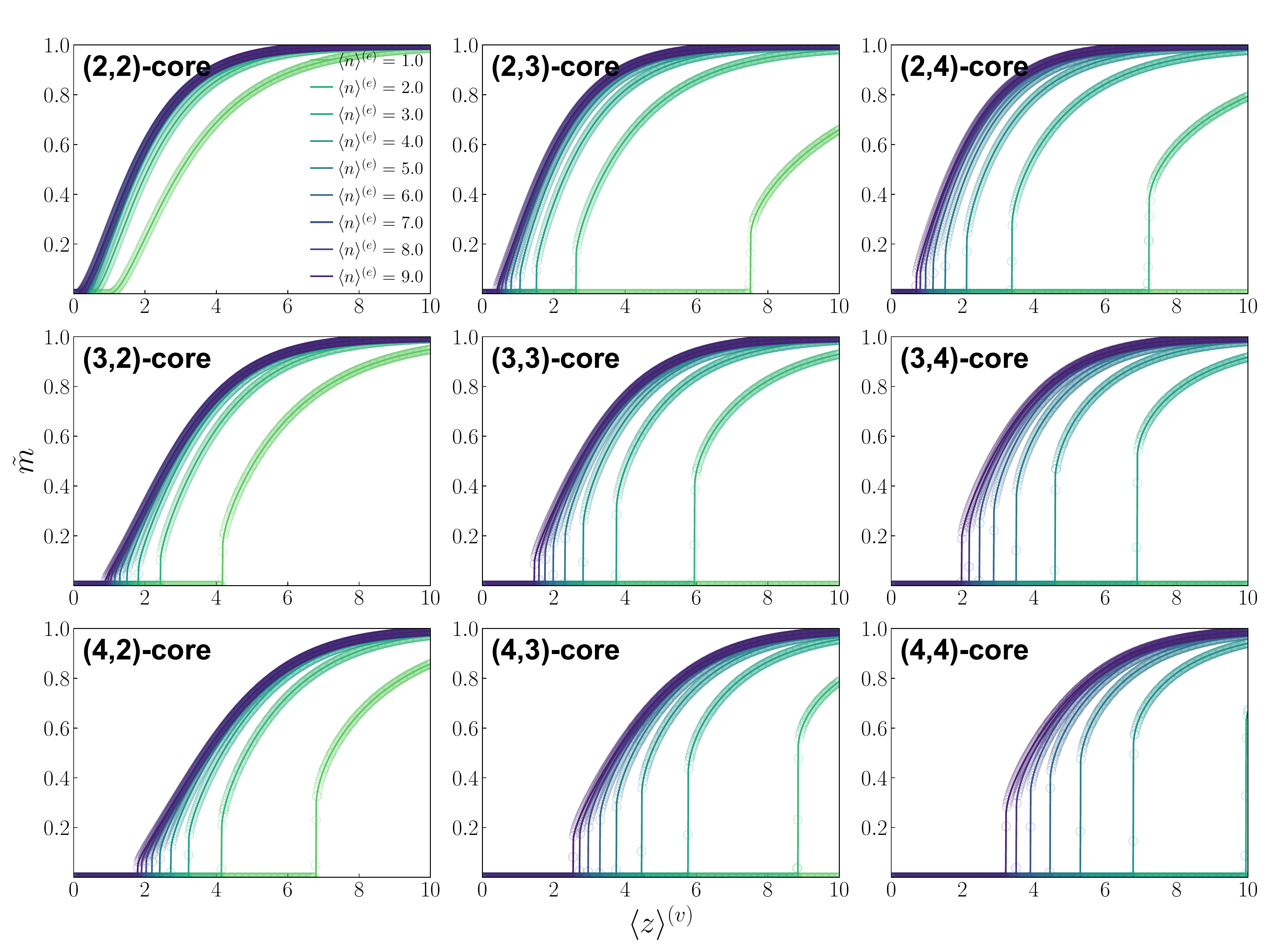}
\caption{\label{fig:Row_slice} Plots of the fraction $\tilde{m}$ of vertices surviving the pruning process as a function of mean vertex degree $\langle z \rangle^{(v)}$ for various $(k,q)$ pairs. The plots depict specific cross-sections of the plots shown in Fig.~\ref{fig:phase_diagram_theory} with respect to the mean hyperedge sizes $\langle n \rangle^{(e)}$. Solid curves represent the numerical solutions of Eqs.~\eqref{eq:exact_evolution_equations} and \eqref{eq:exact_evolution_equations_for_zero}. The Monte Carlo simulation results are exhibited by symbols ($\bigcirc$).}
\end{figure*}

Fig.~\ref{fig:phase_diagram_theory} shows the phase diagram of the $(k,q)$-core percolation transition in random hypergraphs.
Here, we consider the fraction of vertices surviving the pruning process, denoted by $\tilde{m}$. For various mean hyperedge sizes $\langle n\rangle^{(e)}$, we draw the behavior of $\tilde{m}$ as a function of $\langle z \rangle^{(v)}$ in Fig.~\ref{fig:Row_slice}. As the theory predicts, the order parameter exhibits a continuous transition for the cases of $k=2$ and $q=2$ and hybrid transitions for the other cases in Fig.~\ref{fig:phase_diagram_theory}. The order parameter is discontinuous at the transition point. However, it increases continuously for $\langle z \rangle^{(v)} > \langle z \rangle_c^{(v)}$ with the critical behavior as predicted in Eq.~\eqref{eq:order_parameter_hybrid}. Additionally, we observe that the jump size increases with $k$ or $q$. The transition line of $(2,2)$-core percolation in the space of $\langle n \rangle$ and $\langle z \rangle$ is given by Eq.~\eqref{eq:critical22} as~\cite{Newman_2001_PRE_GeneratingFunctions}
\begin{eqnarray}\label{eq:22core_ER_critical_line}
\langle z \rangle^{(v)} \langle n \rangle^{(e)} = 1\,.
\end{eqnarray}
This criterion corresponds to a generalization of the Molloy–Reed criterion of the percolation transition in random networks with pairwise interaction, $\langle k^2\rangle/\langle k \rangle=2$.

\begin{figure}[h!tb]
\includegraphics[width=0.7\linewidth]{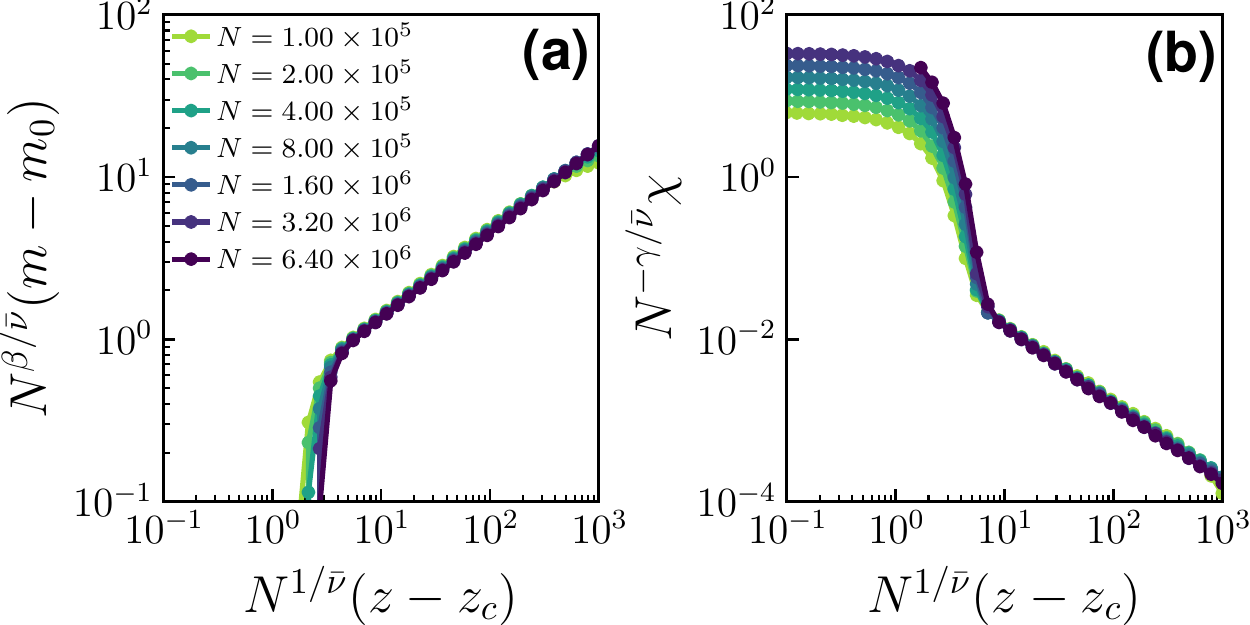}
\caption{\label{fig:FSS_33Core} Finite-size scaling analysis of the order parameter for $(3,3)$-core percolation transition. (a) Scaling plot of the order parameter in the form of $N^{\beta/\bar{\nu}}(m-m_0)$ vs the control parameter in the form of $N^{1/\bar{\nu}}(z-z_c)$. With the selection of $\beta=0.50$ and $\bar{\nu}=2.00$, the data of different system sizes collapse well on a single curve. (b) Scaling plot of the susceptibility in the form of $N^{-\gamma/\bar{\nu}}\chi$ vs the control parameter in the form of $N^{1/\bar{\nu}}(z-z_c)$. With the selection of $\gamma = 1.0$ and $\bar{\nu}=2.00$, the data from different system sizes collapse onto a single curve. The hypergraphs are constructed with the mean hyperedge size $\langle n \rangle^{(e)}=3$. These edges are connected between nodes and hyperedges randomly. For notational simplicity, we drop out the brackets in the notation for $\langle z \rangle^{(v)}$ in (a) and (b).}
\end{figure}

We numerically investigate the critical behavior for $\langle z\rangle^{(v)} > \langle z\rangle_c^{(v)}$ {and given $\langle n\rangle^{(e)}$, using finite size scaling analysis. In Fig.~\ref{fig:FSS_33Core}(a), we plot the order parameter $m$, the fraction of vertices belonging the giant $(k,q)$-core, as a function of mean vertex degree $z$ in the scaling form of $N^{\beta/{\bar \nu}}(m-m_0)$ vs $N^{1/{\bar \nu}}(z-z_c)$. Here, we exploit the simplified notation $z_c = \langle z \rangle_c^{(v)}$. We obtain that the datasets of the order parameters from different system sizes are collapsed onto a single curve with the critical exponent $\beta=0.50$ and $\bar{\nu}=2.00$. This implies that the order parameter behaves as $m-m_0 \sim (z-z_c)^{1/2}$ for $z > z_c$ as in Eq.~\eqref{eq:order_parameter_hybrid}. Thus, the percolation transition is hybrid, exhibiting the features of first- and second-order transitions. The susceptibility $\chi \equiv N(\langle m^2\rangle - \langle m \rangle^2)$, which describes the fluctuations of the order parameter, diverges as $\chi\sim (z-z_c)^{-\gamma}$. The scaling plot of Fig.~\ref{fig:FSS_33Core}(b) decays in a power-law manner asymptotically with slope $-1.00$, indicating that the exponent $\gamma =1.00$. These critical exponents $\beta$, $\gamma$, and $\bar{\nu}$ obtained numerically satisfy the hyperscaling relation $2\beta + \gamma = \bar{\nu}$.

\section{Difference between the $k$-core of graphs and $(k,q)$-core of hypergraphs}\label{sec:realworld}
Here, we discuss the structural difference between the $(k,q)$-core of a hypergraph and the $k$-core of its corresponding graph representation.
Consider a group of $\ell$ vertices forms a hyperedge. In the graph representation, they constitute a clique of $\ell$ fully connected nodes with internal degree $\ell-1$. This clique can be removed completely only by $k$-core decomposition with at least $k\ge \ell$.
In contrast, in the hypergraph representation, vertices or hyperedges associated with such a group can be removed by $(k,q)$-core decomposition for $k$ smaller than $\ell$. This difference becomes more pronounced as $\ell$ becomes larger. Decomposing the graph representation eliminates all the nodes with a degree less than $\ell$, resulting that the remaining structure tending to be over-pruned and grossly fragmented. In contrast, the $(k,q)$-core decomposition with $k<\ell$ can extract the group elements more selectively and distort the other parts of the hypergraph significantly less.

For example, we consider a coauthorship hypergraph comprising authors who published papers citing particular papers~\cite{Watts1998,Barabasi1999,albert_statistical_2002,dorogovtsev_evolution_2002,newman_structure_2003} in the network science field in the early era, during the years 1999--2004~\cite{Yongsun_2021_CHA_HPT}. Here, authors are vertices, and the coauthors of a paper are regarded to form a hyperedge. We selected the network science field because this field is a multidisciplinary field ranging from mathematics and theoretical physics to biological science. Generally, the number of coauthors in the former field is less. In contrast, the number of coauthors in the latter field is routinely more than 10 or even a 100. Therefore, analysis using $k$-core decomposition in graph representation is bound to overvalue the contribution of the researchers from the biological science field.

\begin{figure*}[h!tb]
\includegraphics[width=\linewidth]{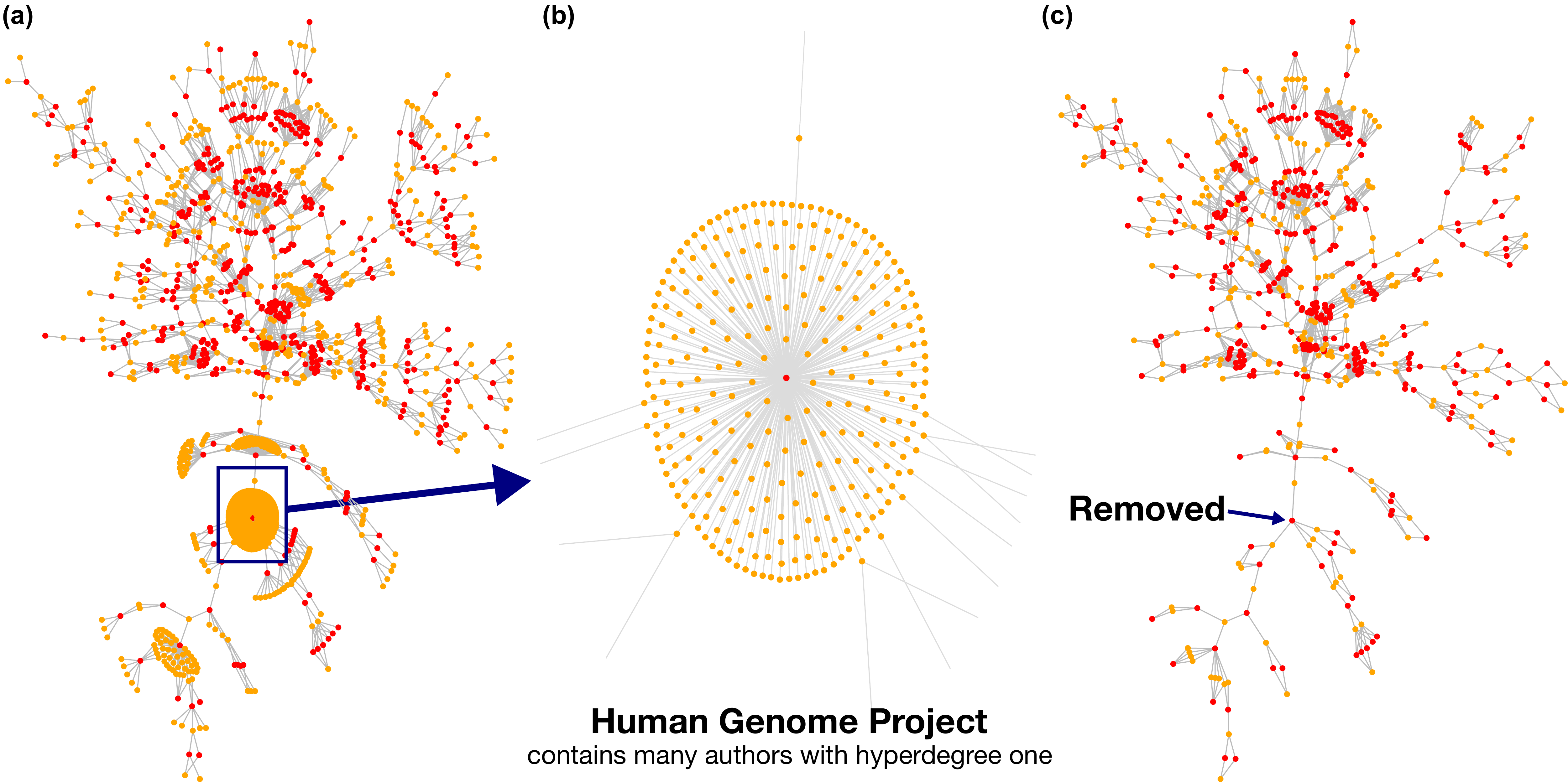}
\caption{\label{fig:coauthorship_remove_genome} Snapshots of the largest connected component (LCC) (a) before and (c) after $(2,2)$-core decomposition of the coauthorship hypergraph of the network science field in December 2004. The dataset contains 3,238 distinct authors and 1,903 articles. In the LCC, only 843 authors and 492 papers are included. For visibility, we utilize bipartite graph representation, where orange (red) nodes represent authors (papers). We focus on the hyperedges (articles) containing many nodes (coauthors), for example, the hyperedge (b). The nodes (orange) are removed from the (2,2)-core decomposition because of their node degrees being 1. However, those nodes remain in $k$-core decomposition in the graph because those nodes have large degrees as they are regarded as being all-to-all connected in the graph. A few such hyperedges (or called groups or modules) of large sizes can be seen in the lower part of the figure; they are removed in the $(2,2)$-core decomposition. The hyperedge shown in (b) is the article of the Human Genome Project ~\cite{Human_Genome_Project_2001_Science}. The $(2,2)$-core in (c) contains 366 authors and 241 articles.}
\end{figure*}

To avoid this overestimation, we apply the $(k,q)$-core decomposition to this coauthorship hypergraph. Figure~\ref{fig:coauthorship_remove_genome} demonstrates that even $(2,2)$-core decomposition can perform well by filtering out authors who contributed to only a single ``mega-project" paper. For example, the coauthorship hypergraph~\cite{Yongsun_2021_CHA_HPT} contains a paper on the Human Genome Project~\cite{Human_Genome_Project_2001_Science} of 274 authors and several papers on yeast genetic interaction~\cite{Yeast_Genetic_Interaction_2004_Science} and Drosophila melanogaster protein interactions~\cite{Protein_Interaction_Map_of_Drosophila_melanogaster_2003_Science}. Many authors of this paper are removed from the $(k,q)$-core decomposition with $k>1$. The $(k,q)$-core of the coauthorship hypergraph indicates that a team composed of $q$ or more coauthors had written more than $k$ papers. This observation implies that the $(k,q)$-core decomposition can be a prototype method for systematically selecting not only large but also active modules. For the coauthorship hypergraph, the core structure is presented in Fig.~\ref{fig:coauthorship_CORE}.

\begin{figure*}[h!tb]
\includegraphics[width=\linewidth]{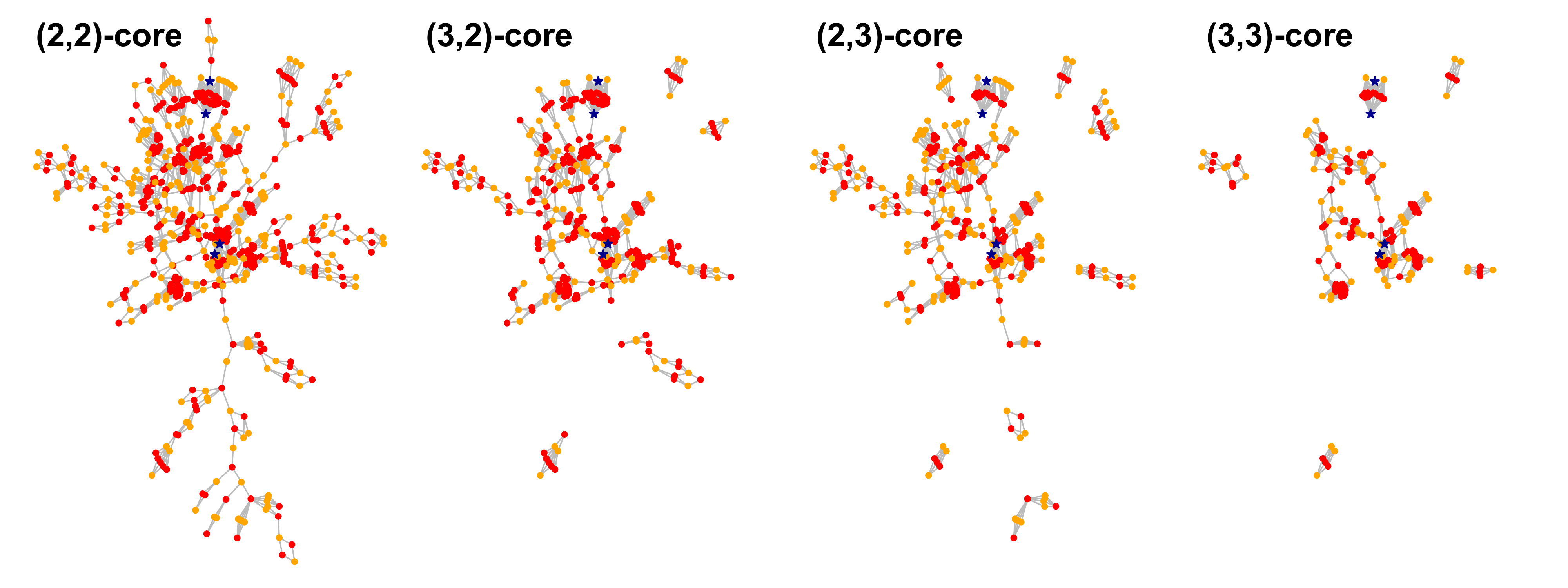}
\caption{\label{fig:coauthorship_CORE}Images of the coauthorship hypergraph decomposed from the largest connected component for various $(k,q)$-core. We denote the authors of the founding papers~\cite{Watts1998,Barabasi1999,albert_statistical_2002,dorogovtsev_evolution_2002,newman_structure_2003} of this coauthorship hypergraph with stars ($\star$). }
\end{figure*}

\begin{figure*}[h!tb]
\centering
\includegraphics[width=0.6\linewidth]{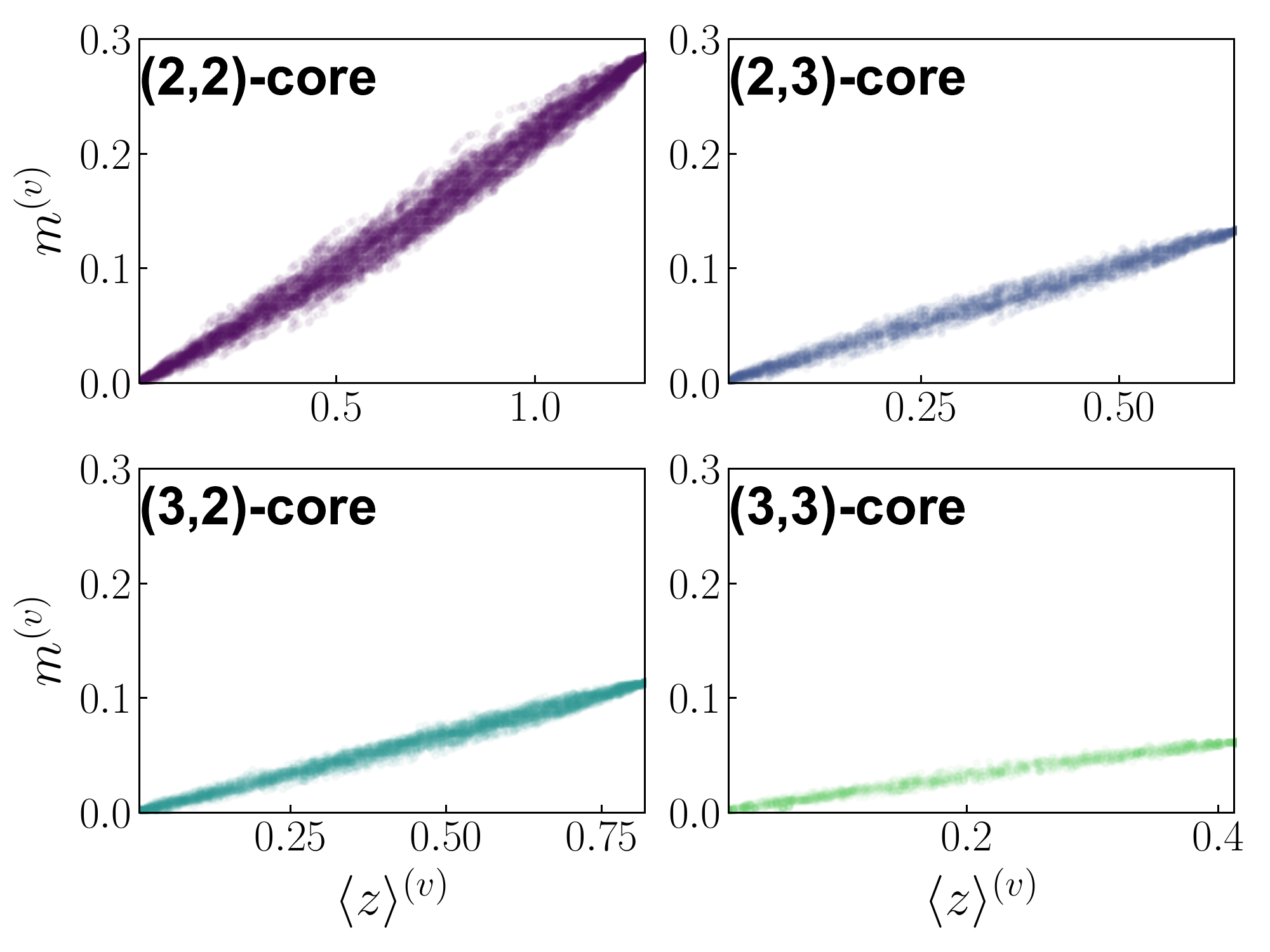}
\caption{\label{fig:coauthorship_core_avalanche} Plot of the order parameter $m^{(v)}$ vs the mean degree $\langle z\rangle^{(v)}$ for the $(k,q)$-core percolation of the coauthorship hypergraph. When a $(k,q)$-core decomposition is implemented, we measure the mean degree $z$ and a randomly selected node is selected and removed. This perturbation induces another pruning process. Afterward, we can get another set of $m^{(v)}$ and $z$. The process is repeated until we reach $m^{(v)}=0$. Here, we take $10^2$ sample average to obtain the data. We obtained that the order parameter increases continuously.}
\end{figure*}

It would be interesting to consider the $(k,q)$-core percolation in the coauthoship hypergraph in which degrees are likely to follow a heavy-tailed distribution, whereas hyperedge sizes follow a Poisson distribution~\cite{Yongsun_2021_CHA_HPT}. We perform numerical simulations by extending the way proposed in Ref.~\cite{Kahng_2016_PRE_kCore}: Once we obtain a $(k,q)$-core of the coauthorship hypergraph, we remove a node randomly selected. This removal can trigger another  cascading pruning process, leading to a new $(k,q)$-core. This process enables us to capture the percolation property from real-world data.
All panels in Fig.~\ref{fig:coauthorship_core_avalanche} show that the order parameter, i.e., the fraction of nodes remaining in the ($k,q$) core decreases to zero continuously. Accordingly, the $(k,q)$-core percolation does not exhibit a hybrid phase transition (see Fig.~\ref{fig:coauthorship_core_avalanche}). Hence, we need further in-depth theoretical study to uncover the cause of this different type of ($k,q$)-core percolation transition. Because the real coauthorship data contain correlations between degrees and/or degree and hyperedge size, the heterogeneity of degree distribution could not be the main matter.

\section{Summary and discussion}\label{discussion}
In this study, we studied $(k,q)$-core decomposition of hypergraphs that encompasses both individual and group units. The evolution equation was derived for uncorrelated hypergraphs to describe the recursive pruning process. The self-consistency equation of the core structure was obtained through the generating functions, and then, the features of $(k,q)$-core percolation transition were elucidated. When $k \ge 3$ or $q \ge 3$ and the second moments of the degree and size distributions are finite, the $(k,q)$-core percolation exhibits a hybrid phase transition. This implies that the order parameter jumps, while critical behavior appears at the transition point. Moreover, the relaxation of the fractions of vertices with degree $z=k-1$ and that of hyperedges of size $n=q-1$ at the transition point exhibits universal behavior $\sim t^{-2}$ for $k\ge 3$ or $q \ge 3$. When $k=q=2$, the novel degree-dependent critical relaxation dynamics was analytically derived and numerically confirmed as $P^{(v)}(z,t)\sim t^{-z}$ and $P^{(e)}(n,t)\sim t^{-n}$ for $z$ and $n \ge 2$, and both decay as $\sim t^{-3}$ for $z=n=1$.  

We showed that the $(k,q)$-core decomposition is more effective to eliminate redundant modular structures than the $k$-core decomposition. 
As a case study, the $(k,q)$-core decomposition was applied to a coauthorship hypergraph to identify the high-impact teams by eliminating modular structures consisting of hyperedges of large sizes and many vertices of small degrees. Indeed, we found that this case study demonstrated the advantage of the $(k,q)$-core decomposition in classifying groups (hyperedges) compared with their properties with respect to the previous $k$-core decomposition. 

Although we have confined our theoretical analysis to hypergraphs with homogeneous distributions in this study, our theory can be generalized to the heterogeneous case. This is due to the convergence of the infinite series, regardless of the divergence of the moments of the degree distribution. Moreover, our method allows for extending to the cases with degree-size correlation. The main limitation of our theory stems from the tree-like assumption, which should be relaxed in future studies to develop more general frameworks. Thus, relaxing this assumption and developing more general frameworks can be promising future researches associated with this work.
\section*{Note added:}
We observed a similar work that appeared in arXiv://arxiv.org/abs/2301.04235 after the completion of this work.
However, the contents of the two papers are different.

\section*{Declaration of Competing Interest}

The authors declare the absence of known competing financial interests or personal relationships that could have appeared to influence the work reported in this study.

\section*{Acknowledgments}
This work was supported by the National Research Foundation in Korea with Grant No. NRF-2014R1A3A2069005 (B.K.), NRF-2019R1A2C1003486 (D.-S.L.), NRF-2020R1A2C2003669 (K.-I.G.), KENTECH Research Grant No. KRG2021-01-007 at Korea Institute of Energy Technology (B.K.), and a KIAS Individual Grant No. CG079901 at Korea Institute for Advanced Study (D.-S.L.).

\section*{Author contributions}
\textbf{Jongshin Lee}: Conceptualization, Methodology, Software, Investigation, Formal analysis, and Writing - Original draft.
\textbf{Kwang-Il Goh}: Conceptualization, Investigation, Formal analysis, Writing - Review and editing, Funding acquisition.
\textbf{Deok-Sun Lee}: Conceptualization, Investigation, Formal analysis, Writing - Review and editing, Funding acquisition.
\textbf{B. Kahng}: Conceptualization, Investigation, Formal analysis, Writing - Review and editing, Funding acquisition, and Supervision.

\section*{Data Availability Statement}
The data are available from the corresponding author upon reasonable request.

\appendix
\counterwithin{figure}{section}
\section{Nomenclature}
This appendix provides a comprehensive summary of the notations used in this paper. Our notations are drawn from the symbols commonly used in the literature on phase transitions and critical phenomena. For instance, the symbol $z$, which denotes degree in the main text, is widely used as a coordination number in the context, and $m$ typically represents the order parameter that distinguishes different phases, such as magnetization of ordered and disordered phase.

The notations used throughout the paper are presented in Table~\ref{tab:TableOfNotation}, whereas Table~\ref{tab:TableOfSubstitutionalvariables} provides a summary of the substitution variables used to avoid multiline equations.

\begin{table}[htbp]\caption{Table of notations}
\begin{center}
\begin{tabular}{r c p{12cm} }
$N$ & & number of vertices (system size $N\equiv |V|$)\\
$M$ & & number of hyperedges (number of interactions $M\equiv |E|$)\\
$L$ & & number of bipartite links in the bipartite graph representation of a hypergraph\\
$z_i$ & & degree of a vertex $v_i$ ($z_i \equiv \deg (v_i)$)\\
$n_i$ & & size of a hyperedge $e_i$ ($n_i \equiv |e_i|$) \\
$\langle z(t)\rangle^{(v)}$ & & mean degree of vertices at pruning time step $t$ ($\equiv \sum_{i\in V}z_i(t)/N$)\\
$\langle n(t)\rangle^{(e)}$ & & average hyperedge size of hyperedges at pruning time step $t$ ($\equiv \sum_{i\in E}n_i(t)/M$)\\
$\langle z\rangle^{(v)}$ & & mean degree of vertices before the pruning, i.e., $\langle z(0)\rangle^{(v)}$ \\
$\langle n\rangle^{(e)}$ & & average hyperedge size of hyperedges before the pruning, i.e., $\langle n(0)\rangle^{(e)}$ \\
$P^{(v)}(z,t)$ & & fraction of vertices having degree $z$ at pruning timestep $t$\\
$P^{(e)}(n,t)$ & & fraction of hyperedges whose size is $n$ at pruning timestep $t$\\
$P^{(v)}(z)$ & & fraction of vertices having degree $z$ before the pruning, i.e., $P^{(v)}(z,0)$\\
$P^{(e)}(n)$ & & fraction of hyperedges whose size is $n$ before the pruning, i.e., $P^{(e)}(n,0)$\\
$R_k^{(v)}(t)$ & & probability of reaching a disqualified vertex at $(k,q)$-pruning timestep $t$ at the end of a bipartite link randomly selected\\
$R_q^{(e)}(t)$ & & probability of reaching a disqualified hyperedge at $(k,q)$-pruning timestep $t$ at the end of a bipartite link randomly selected\\
$\tilde{m}$ & & fraction of vertices survived after the pruning process\\
$m^{(v)}$ & & probability that a vertex belongs to a giant $(k,q)$-core\\
$m^{(e)}$ & & probability that a hyperedge belongs to a giant $(k,q)$-core\\
$R$ & & probability that the end vertex of a randomly-selected bipartite link does not belong to a giant core\\
$H$ & & probability that the end hyperedge of a randomly-selected bipartite link does not belong to a giant core\\
$\Theta_{k,q}^{(v,e)} (x)$ & & self-consistency function of vertices part $\Phi_k^{(v)} \left(\Phi_q^{(e)}(x)\right)$\\
$\Theta_{q,k}^{(e,v)} (x)$ & & self-consistency function of hyperedges part $\Phi_q^{(e)} \left(\Phi_k^{(v)}(x)\right)$\\
$G_1^{(v)}(x)$ & & generating function of the degree distributions $\sum_{z=0}^\infty z P^{(v)}(z) x^{z-1}/\langle z\rangle^{(v)}$\\
$G_1^{(e)}(x)$ & & generating function of the size distributions $\sum_{n=0}^\infty n P^{(e)}(n) x^{n-1}/\langle n\rangle^{(e)}$\\

\end{tabular}
\end{center}
\label{tab:TableOfNotation}
\end{table}

\begin{table}[htbp]\caption{Table of notations for substitutional variables}
\begin{center}
\begin{tabular}{r c p{12cm} }
$T_{z'z}^{(v)}(t)$ & $\equiv$ & $\binom{z'}{z} (1 - R_q^{(e)}(t))^z R_q^{(e)}(t)^{z' -z}$ \\
$T_{n'n}^{(e)}(t)$ & $\equiv$ & $\binom{n'}{n} (1 - R_k^{(v)}(t))^n R_k^{(v)}(t)^{n' -n}$ \\
$\Phi_k^{(v)}(x)$ & $\equiv$ & $\sum_{r=0}^{k-2} \sum_{z=r}^\infty \frac{(z+1) P^{(v)}(z+1)}{ \langle z\rangle^{(v)}} \binom{z}{r} (1-x)^r x^{z-r}$\\
$\Phi_q^{(e)}(x)$ & $\equiv$ & $\sum_{s=0}^{q-2} \sum_{n=s}^\infty \frac{(n+1) P^{(e)}(n+1)}{ \langle n\rangle^{(e)}} \binom{n}{s} (1-x)^s x^{n-s}$\\
$\Delta$ & $ \equiv $ & ${\frac{\langle z(z-1)\rangle^{(v)}} {\langle z\rangle^{(v)}}}{\frac{\langle n(n-1)\rangle^{(e)}}{\langle n\rangle^{(e)}}}-1$\\
\end{tabular}
\end{center}
\label{tab:TableOfSubstitutionalvariables}
\end{table}

\section{Analysis of the critical relaxation of $(2,2)$-core pruning}\label{Appendix:Analytic_calculation_for_22core}
This section aims to derive the temporal behaviors of $P^{(v)}({m},{t})$ and $P^{(e)}({n},{t})$ during the $(2,2)$-core pruning process.
To achieve our goal, we utilize the noncrossing approximation, introduced in Ref.~\cite{Baxter_PRX_kCore}, that the probability to reach a disqualified vertex or a hyperedge along a link is small and the branching process can approximate the critical pruning process. In this approximation, the fractions of the vertices of degree $z < k - 1$ and hyperedges of size $ n < q-1 $ contribute little to the branching process. However, the vertices of degree $z = k - 1$ and hyperedges of size $n= q-1$ contribute dominantly. Thus, Eqs.~\eqref{eq:exact_removal_prob} are replaced by the following equations,
\begin{eqnarray}
    \begin{aligned}
        R^{(v)}(t) = \frac{(k-1)P^{(v)}({k-1},{t})}{\langle z(t)\rangle^{(v)}}\,,\quad R^{(e)}(t) = \frac{(q-1)P^{(e)}({q-1},{t})}{\langle n(t)\rangle^{(v)}}\,.
    \end{aligned}\label{eq:removal_prob}
\end{eqnarray}

Expanding the exact equations Eqs.~\eqref{eq:exact_evolution_equations} up to the first order of $R^{(v)}(t)$ and $R^{(e)}(t)$ gives the followings:
\begin{eqnarray}\label{eq:noncrossing_dynamics_z}
    \begin{aligned}
    &P^{(v)}({z},{t+1}) = P^{(v)}({z},{t})- R^{(e)}(t)z P^{(v)}({z},{t})+ R^{(e)}(t)(z+1)P^{(v)}({z+1},{t})\,, &\text{ for }z\ge k\,,\\
    &P^{(e)}({n},{t+1}) = P^{(e)}({n},{t})- R^{(v)}(t)n P^{(e)}({n},{t})+ R^{(v)}(t)(n+1)P^{(e)}({n+1},{t})\,, &\text{  for }n\ge q\,,
    \end{aligned}
\end{eqnarray}
\begin{eqnarray}\label{eq:noncrossing_dynamics_k-1}
    \begin{aligned}
    &P^{(v)}({k-1},{t+1}) = R^{(e)}(t) k P^{(v)}({k},{t})\,,\\
    &P^{(e)}({q-1},{t+1}) = R^{(v)}(t) q P^{(e)}({q},{t})\,,
    \end{aligned}
\end{eqnarray}
\begin{eqnarray}\label{eq:noncrossing_dynamics_0}
    \begin{aligned}
    &P^{(v)}({0},{t+1}) = P^{(v)}({0},{t})+P^{(v)}({k-1},{t})\,, \\
    &P^{(e)}({0},{t+1}) = P^{(e)}({0},{t})+P^{(v)}({q-1},{t})\,.
    \end{aligned}
\end{eqnarray}
Then, the mean vertex degree and the mean hyperedge size become as follows:
\begin{eqnarray}\label{eq:noncrossing_dynamics_meandeg}
    \begin{aligned}
    \langle z(t)\rangle^{(v)} = (k-1)P^{(v)}({k-1},{t})+\sum_{z\ge k} z P^{(v)}({z},{t})\,, \qquad \langle n(t)\rangle^{(e)} = (q-1)P^{(e)}({q-1},{t})+\sum_{n\ge q} n P^{(e)}({n},{t})\,.
    \end{aligned}
\end{eqnarray}

We focus on the case of $k=q=2$. The fractions of the primary disqualified vertices and hyperedges in Eqs.~\eqref{eq:noncrossing_dynamics_k-1} evolve with time as
\begin{eqnarray}
    \begin{aligned}
        P^{(v)}({1},{t+1}) =& 2 R^{(e)}(t) P^{(v)}({2},{t})\,,\\
        P^{(e)}({1},{t+1}) =& 2 R^{(v)}(t) P^{(e)}({2},{t})\,.
    \end{aligned}
\end{eqnarray}
At the critical point, we expect that the fractions are practically preserved as in the critical branching process,
\begin{eqnarray}
    \begin{aligned}
        P^{(v)}({1}{t+1})\simeq&P^{(v)}({1},{t})\,,\\
        P^{(e)}({1},{t+1})\simeq&P^{(e)}({1},{t})\,.
    \end{aligned}
\end{eqnarray}
which are inserted into the product of the two fractions
\begin{eqnarray}
    \begin{aligned}
        P^{(v)}&(1,t+1)P^{(e)}(1,t+1)\\ &= 4\frac{P^{(v)}(1,t)P^{(e)}(1,t)}{\langle z\rangle^{(v)}\langle n\rangle^{(e)}}P^{(v)}(2,t)P^{(e)}(2,t)\,,
    \end{aligned}
\end{eqnarray}
to yield
\begin{eqnarray}
    \langle z \rangle^{(v)}\langle n\rangle^{(e)} = 4P^{(v)}(2,t)P^{(e)}(2,t)\,.
\end{eqnarray}

Since
\begin{align}
    \langle z \rangle^{(v)}&\langle n\rangle^{(e)} = (P^{(v)}(1,t)+2P^{(v)}(2,t)+3P^{(v)}(3,t)+\cdots)\nonumber\\
        &\times(P^{(e)}(1,t)+2P^{(e)}(2,t)+3P^{(e)}(3,t)+\cdots)\,,
\end{align}
we can consider that
\begin{eqnarray}
    \begin{aligned}
        P^{(v)}({2},{t})\gg& P^{(v)}({z},{t}) \text{ for all }z>0 \text{ but }z=2 \,,\\
        P^{(e)}({2},{t})\gg& P^{(e)}({n},{t}) \text{ for all }n>0 \text{ but }n=2 \,.
    \end{aligned}\label{eq:assumptions_for_calculations}
\end{eqnarray}
The assumptions are valid as shown by the numerical solutions to the exact time evolution equations in Fig.~\ref{fig:2core_power_law_degree_evolution}. In the case of $z>2$, we expect that the percolation cluster at the critical point of the ER network is similar to the tree structure.

We remark that when $k=q$ and $\langle z \rangle^{(v)} = \langle n \rangle^{(e)}$, $P^{(v)}({z},{t})$ and $P^{(e)}({n},{t})$ becomes the same. Therefore, the above derivation can be simplified. In this case, the oscillating pattern of $\langle z \rangle^{(v)}$ disappears.

Since we are currently considering $(2,2)$-core pruning, we proceed with calculating at $\langle z \rangle^{(v)} = \langle n \rangle^{(e)} = 1$ among the numerous combinations of critical lines in Eq.~\eqref{eq:22core_ER_critical_line}. This case perfectly matches the two-core dynamics in the graph.
Considering that $P^{(v)}(z,t)\gg P^{(v)}(z+1,t)$ and $P^{(e)}(n,t)\gg P^{(e)}(n+1,t)$ for all $z>2$ and $n>2$, which will be shown to be self-consistent and will be confirmed numerically by Fig.~\ref{fig:2core_power_law_degree_evolution}, we obtain the following equations in continuous time from Eqs.~(\ref{eq:noncrossing_dynamics_z}--\ref{eq:noncrossing_dynamics_meandeg}):

\begin{align}
    \frac{\partial P^{(v)}({z},{t})}{\partial t} =& -\frac{zP^{(v)}({z},{t})}{\langle z(t) \rangle^{(v)}} P^{(v)}({1},{t})\,,\text{ }\forall z\ge 2\,,\label{eq:appendix_meanfield_a}\\
    \frac{\partial P^{(v)}({1},{t})}{\partial t} =& \left[\frac{2P^{(v)}({2},{t})}{\langle z(t)\rangle^{(v)} }-1\right] P^{(v)}({1},{t})\,, \label{eq:appendix_meanfield_b}\\
    \frac{\partial P^{(v)}({0},{t})}{\partial t} =& P^{(v)}({1},{t})\,,\label{eq:appendix_meanfield_c}\\
    \langle z(t)\rangle^{(v)} =& P^{(v)}({1},{t})+2P^{(v)}({2},{t})+3P^{(v)}({3},{t})\,.\label{eq:appendix_meanfield_d}
\end{align}

The equations can be solved by an ansatz $P^{(v)}({z},{t})\simeq a_z t^{-b_z}$. Eq.~\eqref{eq:removal_prob} is then represented as
\begin{eqnarray}
    \begin{aligned}
        R^{(v)}(t) \simeq \frac{P^{(v)}({1},{t})}{\langle z(t)\rangle^{(v)}} = \frac{P^{(v)}({1},{t})}{2P^{(v)}({2},{t})}\simeq \frac{a_1}{2a_2}t^{-(b_1-b_2)}\,.
    \end{aligned}
\end{eqnarray}
Eq.~\eqref{eq:appendix_meanfield_a} for $z\ge 2$ becomes

\begin{eqnarray}
    -a_zb_z t^{-b_z-1} = -\frac{z a_1a_z}{2 a_2} t^{-b_z -(b_1-b_2)}\,;
\end{eqnarray}
thus, we can get the following recurrence relation for $b_z$:
\begin{eqnarray}\label{eq:appendix_recurrence}
    b_1 - b_2 = 1 \text{, and }b_z = \frac{a_1}{2a_2} z \text{ for }z\ge 2\,.
\end{eqnarray}

Moreover, Eq.~\eqref{eq:appendix_meanfield_b} for $z=1$ becomes
\begin{eqnarray*}
    -a_1b_1t^{-b_1-1} = \left(-\frac{a_1}{2a_2}t^{-(b_1-b_2)} - \frac{3a_3}{2a_2}t^{-(b_3-b_2)}\right)a_1t_1^{-b_1}\,,
\end{eqnarray*}
so that $ b_3 - b_2 = 1$. Therefore, using the $b_z$ of the Eq.~\eqref{eq:appendix_recurrence}, we get
\begin{eqnarray}
    \frac{a_1}{2a_2}= 1\,.
\end{eqnarray}
Finally, in summary, we get
\begin{eqnarray}
    b_z = \begin{cases}
        z & \text{for }z\ge 2\,,\\
        3 & \text{for }z=1\,.
    \end{cases}
\end{eqnarray}
This means that
\begin{eqnarray}
        P^{(v)}({z},{t})\simeq& \begin{cases}
            a_1 t^{-3} &\text{for }z=1\,,\\
            a_z t^{-z} &\text{for }z\ge 2\,.
        \end{cases}
\end{eqnarray}
Therefore, we obtain the dynamic exponent $\sigma =3$ for the $(2,2)$-core pruning process. This result is confirmed by numerical computations, as shown in Fig.~\ref{fig:2core_power_law_degree_evolution}. If $\langle z \rangle^{(v)}\neq \langle n \rangle^{(e)}$, $P^{(v)}(z,t)$ is not the same as $P^{(e)}(n,t)$. However, their decay patterns are the same.

\section{Behavior of self-consistency equations for ER random hypergraphs}\label{Appendix:Self_Consistency}

\begin{figure*}[h!tb]
\includegraphics[width=\linewidth]{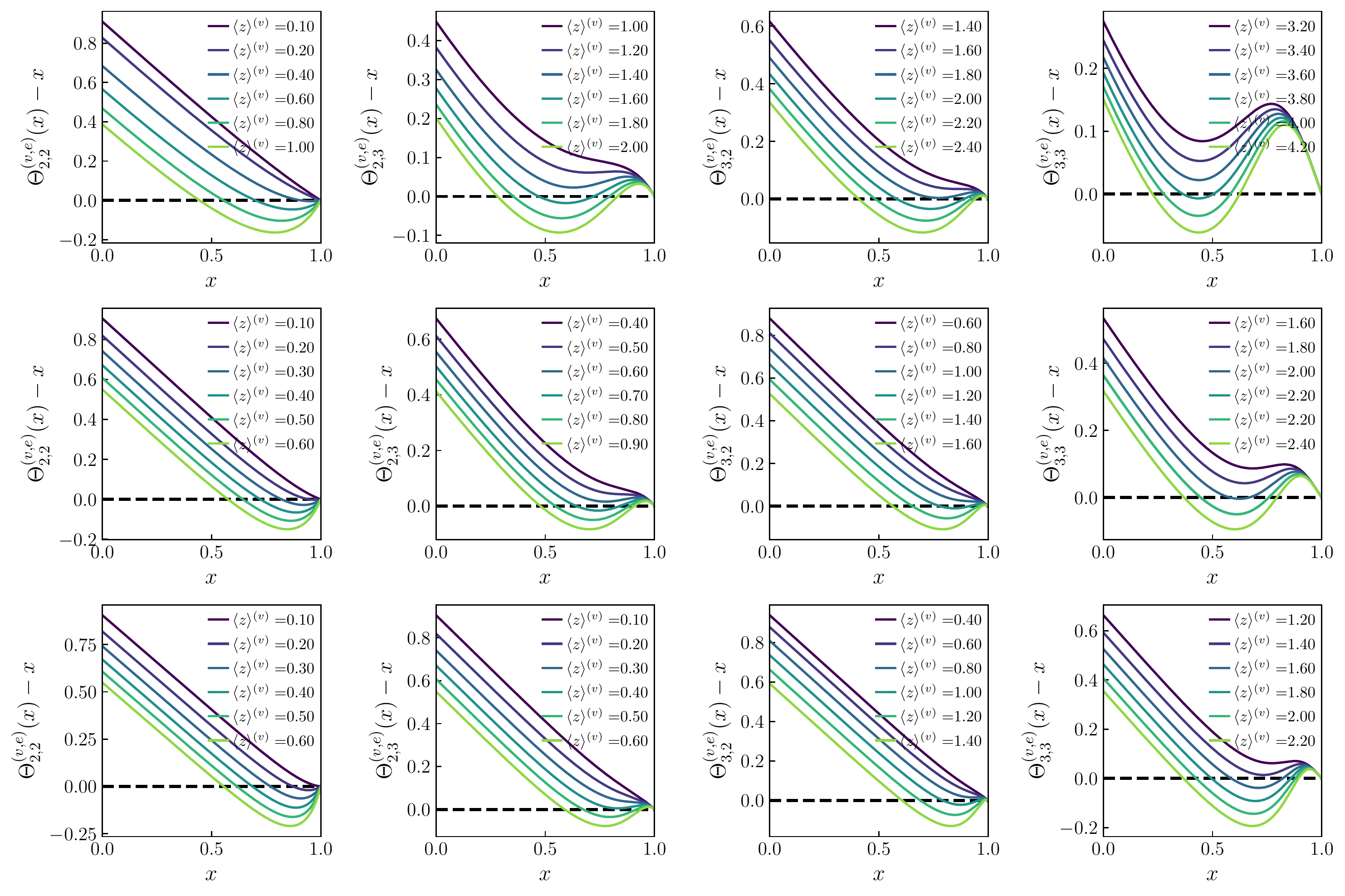}
\caption{\label{fig:self_consistency} Behavior of the self-consistency functions $\Theta_{k,q}^{(v,e)}(x)-x$ for various $(k,q)$ cores: Each column from left to right is for $(k,q)=(2,2)$, $(2,3)$, $(3,2)$, and $(3,3)$ cores, respectively. Each row from the top to bottom is for the mean hyperedge size $\langle n\rangle^{(e)}=3.0$, $6.0$, and $9.0$, respectively. Different mean node degrees $\langle z \rangle^{(v)}$ are denoted as shown in the legend of each panel. }
\end{figure*}
In this section, we numerically investigate the behavior of the self-consistency equation for the random hypergraph. As we show in the main text, the behavior of $\Theta_{k,q}^{(v,e)}(x)-x$ can determine the type of phase transition, threshold, and jump size. Figure~\ref{fig:self_consistency} confirms the behavior of $\Theta_{k,q}^{(v,e)}$ for various $k$ and $q$.


\begin{thebibliography}{10}
\expandafter\ifx\csname url\endcsname\relax
  \def\url#1{\texttt{#1}}\fi
\expandafter\ifx\csname urlprefix\endcsname\relax\def\urlprefix{URL }\fi
\expandafter\ifx\csname href\endcsname\relax
  \def\href#1#2{#2} \def\path#1{#1}\fi

\bibitem{Battiston_2020_HigherOrderReview}
F.~Battiston, G.~Cencetti, I.~Iacopini, V.~Latora, M.~Lucas, A.~Patania, J.-G. Young, G.~Petri, \href{https://www.sciencedirect.com/science/article/pii/S0370157320302489}{Networks beyond pairwise interactions: Structure and dynamics}, Phys. Rep. 874 (2020) 1--92.
\newblock \href {https://doi.org/https://doi.org/10.1016/j.physrep.2020.05.004} {\path{doi:https://doi.org/10.1016/j.physrep.2020.05.004}}.
\newline\urlprefix\url{https://www.sciencedirect.com/science/article/pii/S0370157320302489}

\bibitem{Battiston_2021_HigherOrder_NatPhys}
F.~Battiston, E.~Amico, A.~Barrat, G.~Bianconi, G.~Ferraz~de Arruda, B.~Franceschiello, I.~Iacopini, S.~K{\'e}fi, V.~Latora, Y.~Moreno, M.~M. Murray, T.~P. Peixoto, F.~Vaccarino, G.~Petri, \href{https://doi.org/10.1038/s41567-021-01371-4}{The physics of higher-order interactions in complex systems}, Nat. Phys. 17~(10) (2021) 1093--1098.
\newblock \href {https://doi.org/10.1038/s41567-021-01371-4} {\path{doi:10.1038/s41567-021-01371-4}}.
\newline\urlprefix\url{https://doi.org/10.1038/s41567-021-01371-4}

\bibitem{Torres_2021_SIAMreview}
L.~Torres, A.~S. Blevins, D.~Bassett, T.~Eliassi-Rad, \href{https://doi.org/10.1137/20M1355896}{The why, how, and when of representations for complex systems}, SIAM Rev. 63~(3) (2021) 435--485.
\newblock \href {https://doi.org/10.1137/20M1355896} {\path{doi:10.1137/20M1355896}}.
\newline\urlprefix\url{https://doi.org/10.1137/20M1355896}

\bibitem{Petri_2018_SimplicialActivityDrivenModel}
G.~Petri, A.~Barrat, \href{https://link.aps.org/doi/10.1103/PhysRevLett.121.228301}{Simplicial activity driven model}, Phys. Rev. Lett. 121 (2018) 228301.
\newblock \href {https://doi.org/10.1103/PhysRevLett.121.228301} {\path{doi:10.1103/PhysRevLett.121.228301}}.
\newline\urlprefix\url{https://link.aps.org/doi/10.1103/PhysRevLett.121.228301}

\bibitem{Iacopini_2019_SimplicialSocialContagion}
I.~Iacopini, G.~Petri, A.~Barrat, V.~Latora, \href{https://doi.org/10.1038/s41467-019-10431-6}{Simplicial models of social contagion}, Nat. Commun. 10~(1) (2019) 2485.
\newblock \href {https://doi.org/10.1038/s41467-019-10431-6} {\path{doi:10.1038/s41467-019-10431-6}}.
\newline\urlprefix\url{https://doi.org/10.1038/s41467-019-10431-6}

\bibitem{Jhun_2019_SISinHypergraph}
B.~Jhun, M.~Jo, B.~Kahng, \href{https://doi.org/10.1088/1742-5468/ab5367}{Simplicial {SIS} model in scale-free uniform hypergraph}, J. Stat. Mech.: Theory Exp. 2019~(12) (2019) 123207.
\newblock \href {https://doi.org/10.1088/1742-5468/ab5367} {\path{doi:10.1088/1742-5468/ab5367}}.
\newline\urlprefix\url{https://doi.org/10.1088/1742-5468/ab5367}

\bibitem{Arenas_2020_PRR_EpidemicsInSimplicialComplex}
J.~T. Matamalas, S.~G\'omez, A.~Arenas, \href{https://link.aps.org/doi/10.1103/PhysRevResearch.2.012049}{Abrupt phase transition of epidemic spreading in simplicial complexes}, Phys. Rev. Research 2 (2020) 012049.
\newblock \href {https://doi.org/10.1103/PhysRevResearch.2.012049} {\path{doi:10.1103/PhysRevResearch.2.012049}}.
\newline\urlprefix\url{https://link.aps.org/doi/10.1103/PhysRevResearch.2.012049}

\bibitem{Carletti_2020_PRE_RWonHypergraphs}
T.~Carletti, F.~Battiston, G.~Cencetti, D.~Fanelli, \href{https://link.aps.org/doi/10.1103/PhysRevE.101.022308}{Random walks on hypergraphs}, Phys. Rev. E 101 (2020) 022308.
\newblock \href {https://doi.org/10.1103/PhysRevE.101.022308} {\path{doi:10.1103/PhysRevE.101.022308}}.
\newline\urlprefix\url{https://link.aps.org/doi/10.1103/PhysRevE.101.022308}

\bibitem{Arenas_2019_PRL_OscillatorSimplexes}
P.~S. Skardal, A.~Arenas, \href{https://link.aps.org/doi/10.1103/PhysRevLett.122.248301}{Abrupt desynchronization and extensive multistability in globally coupled oscillator simplexes}, Phys. Rev. Lett. 122 (2019) 248301.
\newblock \href {https://doi.org/10.1103/PhysRevLett.122.248301} {\path{doi:10.1103/PhysRevLett.122.248301}}.
\newline\urlprefix\url{https://link.aps.org/doi/10.1103/PhysRevLett.122.248301}

\bibitem{Skardal_2020_CommPhys_HigherOrderSync}
P.~S. Skardal, A.~Arenas, \href{https://doi.org/10.1038/s42005-020-00485-0}{Higher order interactions in complex networks of phase oscillators promote abrupt synchronization switching}, Commun. Phys. 3~(1) (2020) 218.
\newblock \href {https://doi.org/10.1038/s42005-020-00485-0} {\path{doi:10.1038/s42005-020-00485-0}}.
\newline\urlprefix\url{https://doi.org/10.1038/s42005-020-00485-0}

\bibitem{Bianconi_2020_PRL_SimplicialKuramoto}
A.~P. Mill\'an, J.~J. Torres, G.~Bianconi, \href{https://link.aps.org/doi/10.1103/PhysRevLett.124.218301}{Explosive higher-order kuramoto dynamics on simplicial complexes}, Phys. Rev. Lett. 124 (2020) 218301.
\newblock \href {https://doi.org/10.1103/PhysRevLett.124.218301} {\path{doi:10.1103/PhysRevLett.124.218301}}.
\newline\urlprefix\url{https://link.aps.org/doi/10.1103/PhysRevLett.124.218301}

\bibitem{Mulars_2020_PRE_MasterStabilityOnHypergraphs}
R.~Mulas, C.~Kuehn, J.~Jost, \href{https://link.aps.org/doi/10.1103/PhysRevE.101.062313}{Coupled dynamics on hypergraphs: Master stability of steady states and synchronization}, Phys. Rev. E 101 (2020) 062313.
\newblock \href {https://doi.org/10.1103/PhysRevE.101.062313} {\path{doi:10.1103/PhysRevE.101.062313}}.
\newline\urlprefix\url{https://link.aps.org/doi/10.1103/PhysRevE.101.062313}

\bibitem{Latora_2021_NHB_EvolutionOfSocialNetworks}
U.~Alvarez-Rodriguez, F.~Battiston, G.~F. de~Arruda, Y.~Moreno, M.~Perc, V.~Latora, \href{https://doi.org/10.1038/s41562-020-01024-1}{Evolutionary dynamics of higher-order interactions in social networks}, Nat. Hum. Behav. 5~(5) (2021) 586--595.
\newblock \href {https://doi.org/10.1038/s41562-020-01024-1} {\path{doi:10.1038/s41562-020-01024-1}}.
\newline\urlprefix\url{https://doi.org/10.1038/s41562-020-01024-1}

\bibitem{Jongshin_2021_CHA_BC}
J.~Lee, Y.~Lee, S.~M. Oh, B.~Kahng, \href{https://doi.org/10.1063/5.0056683}{Betweenness centrality of teams in social networks}, Chaos 31~(6) (2021) 061108.
\newblock \href {https://doi.org/10.1063/5.0056683} {\path{doi:10.1063/5.0056683}}.
\newline\urlprefix\url{https://doi.org/10.1063/5.0056683}

\bibitem{Giusti_2016_Neuroscience_Simplex}
C.~Giusti, R.~Ghrist, D.~S. Bassett, \href{https://doi.org/10.1007/s10827-016-0608-6}{Two's company, three (or more) is a simplex}, J. Comput. Neurosci. 41~(1) (2016) 1--14.
\newblock \href {https://doi.org/10.1007/s10827-016-0608-6} {\path{doi:10.1007/s10827-016-0608-6}}.
\newline\urlprefix\url{https://doi.org/10.1007/s10827-016-0608-6}

\bibitem{Jost_2019_HypergraphChemicalReaction}
J.~Jost, R.~Mulas, \href{https://www.sciencedirect.com/science/article/pii/S0001870819302671}{Hypergraph laplace operators for chemical reaction networks}, Adv. Math. 351 (2019) 870--896.
\newblock \href {https://doi.org/https://doi.org/10.1016/j.aim.2019.05.025} {\path{doi:https://doi.org/10.1016/j.aim.2019.05.025}}.
\newline\urlprefix\url{https://www.sciencedirect.com/science/article/pii/S0001870819302671}

\bibitem{Bianconi_PRE2021_Higher-Order_Multiplex_Percolation}
H.~Sun, G.~Bianconi, \href{https://link.aps.org/doi/10.1103/PhysRevE.104.034306}{Higher-order percolation processes on multiplex hypergraphs}, Phys. Rev. E 104 (2021) 034306.
\newblock \href {https://doi.org/10.1103/PhysRevE.104.034306} {\path{doi:10.1103/PhysRevE.104.034306}}.
\newline\urlprefix\url{https://link.aps.org/doi/10.1103/PhysRevE.104.034306}

\bibitem{Coutinho_PRL2020_GRL}
B.~C. Coutinho, A.-K. Wu, H.-J. Zhou, Y.-Y. Liu, \href{https://link.aps.org/doi/10.1103/PhysRevLett.124.248301}{Covering problems and core percolations on hypergraphs}, Phys. Rev. Lett. 124 (2020) 248301.
\newblock \href {https://doi.org/10.1103/PhysRevLett.124.248301} {\path{doi:10.1103/PhysRevLett.124.248301}}.
\newline\urlprefix\url{https://link.aps.org/doi/10.1103/PhysRevLett.124.248301}

\bibitem{Newman_2018_Networks}
M.~Newman, Networks, Oxford University Press, 2018.

\bibitem{Seidman_1983_k_core}
S.~B. Seidman, \href{https://www.sciencedirect.com/science/article/pii/037887338390028X}{Network structure and minimum degree}, Soc. Networks 5~(3) (1983) 269--287.
\newblock \href {https://doi.org/https://doi.org/10.1016/0378-8733(83)90028-X} {\path{doi:https://doi.org/10.1016/0378-8733(83)90028-X}}.
\newline\urlprefix\url{https://www.sciencedirect.com/science/article/pii/037887338390028X}

\bibitem{Kong_2019_PhysRep_kCore}
Y.-X. Kong, G.-Y. Shi, R.-J. Wu, Y.-C. Zhang, \href{https://www.sciencedirect.com/science/article/pii/S037015731930328X}{k-core: Theories and applications}, Phys. Rep. 832 (2019) 1--32, k-core: Theories and Applications.
\newblock \href {https://doi.org/https://doi.org/10.1016/j.physrep.2019.10.004} {\path{doi:https://doi.org/10.1016/j.physrep.2019.10.004}}.
\newline\urlprefix\url{https://www.sciencedirect.com/science/article/pii/S037015731930328X}

\bibitem{Stanley_NatPhys2010_influential_spreaders}
M.~Kitsak, L.~K. Gallos, S.~Havlin, F.~Liljeros, L.~Muchnik, H.~E. Stanley, H.~A. Makse, \href{https://doi.org/10.1038/nphys1746}{Identification of influential spreaders in complex networks}, Nat. Phys. 6~(11) (2010) 888--893.
\newblock \href {https://doi.org/10.1038/nphys1746} {\path{doi:10.1038/nphys1746}}.
\newline\urlprefix\url{https://doi.org/10.1038/nphys1746}

\bibitem{Motter_Science2017_Vulnerable_set_in_Power_grid}
Y.~Yang, T.~Nishikawa, A.~E. Motter, \href{https://www.science.org/doi/abs/10.1126/science.aan3184}{Small vulnerable sets determine large network cascades in power grids}, Science 358~(6365) (2017) eaan3184.
\newblock \href {https://doi.org/10.1126/science.aan3184} {\path{doi:10.1126/science.aan3184}}.
\newline\urlprefix\url{https://www.science.org/doi/abs/10.1126/science.aan3184}

\bibitem{Wuchty_2005_Protein}
S.~Wuchty, E.~Almaas, \href{https://analyticalsciencejournals.onlinelibrary.wiley.com/doi/abs/10.1002/pmic.200400962}{Peeling the yeast protein network}, PROTEOMICS 5~(2) (2005) 444--449.
\newblock \href {https://doi.org/https://doi.org/10.1002/pmic.200400962} {\path{doi:https://doi.org/10.1002/pmic.200400962}}.
\newline\urlprefix\url{https://analyticalsciencejournals.onlinelibrary.wiley.com/doi/abs/10.1002/pmic.200400962}

\bibitem{KLIMEK_2009_Evolution}
P.~Klimek, S.~Thurner, R.~Hanel, \href{https://www.sciencedirect.com/science/article/pii/S0022519308004979}{Pruning the tree of life: k-core percolation as selection mechanism}, J. Theor. Biol. 256~(1) (2009) 142--146.
\newblock \href {https://doi.org/https://doi.org/10.1016/j.jtbi.2008.09.030} {\path{doi:https://doi.org/10.1016/j.jtbi.2008.09.030}}.
\newline\urlprefix\url{https://www.sciencedirect.com/science/article/pii/S0022519308004979}

\bibitem{Schwarz_2006_EPL_Jamming}
J.~M. Schwarz, A.~J. Liu, L.~Q. Chayes, \href{https://doi.org/10.1209/epl/i2005-10421-7}{The onset of jamming as the sudden emergence of an infinite $k$-core cluster}, Europhys. Lett. 73~(4) (2006) 560--566.
\newblock \href {https://doi.org/10.1209/epl/i2005-10421-7} {\path{doi:10.1209/epl/i2005-10421-7}}.
\newline\urlprefix\url{https://doi.org/10.1209/epl/i2005-10421-7}

\bibitem{Dorogovtsev_2006_PRL_kCore}
S.~N. Dorogovtsev, A.~V. Goltsev, J.~F.~F. Mendes, \href{https://link.aps.org/doi/10.1103/PhysRevLett.96.040601}{$k$-core organization of complex networks}, Phys. Rev. Lett. 96 (2006) 040601.
\newblock \href {https://doi.org/10.1103/PhysRevLett.96.040601} {\path{doi:10.1103/PhysRevLett.96.040601}}.
\newline\urlprefix\url{https://link.aps.org/doi/10.1103/PhysRevLett.96.040601}

\bibitem{Goltsev_2006_PRE_kCore}
A.~V. Goltsev, S.~N. Dorogovtsev, J.~F.~F. Mendes, \href{https://link.aps.org/doi/10.1103/PhysRevE.73.056101}{$k$-core (bootstrap) percolation on complex networks: Critical phenomena and nonlocal effects}, Phys. Rev. E 73 (2006) 056101.
\newblock \href {https://doi.org/10.1103/PhysRevE.73.056101} {\path{doi:10.1103/PhysRevE.73.056101}}.
\newline\urlprefix\url{https://link.aps.org/doi/10.1103/PhysRevE.73.056101}

\bibitem{Dorogovtsev_2006_PhysicaD_kcore}
S.~Dorogovtsev, A.~Goltsev, J.~Mendes, \href{https://www.sciencedirect.com/science/article/pii/S0167278906003617}{k-core architecture and k-core percolation on complex networks}, Physica D: Nonlinear Phenomena 224~(1) (2006) 7--19, dynamics on Complex Networks and Applications.
\newblock \href {https://doi.org/https://doi.org/10.1016/j.physd.2006.09.027} {\path{doi:https://doi.org/10.1016/j.physd.2006.09.027}}.
\newline\urlprefix\url{https://www.sciencedirect.com/science/article/pii/S0167278906003617}

\bibitem{Baxter_2011_PRE_Bootstrap_vs_Kcore}
G.~J. Baxter, S.~N. Dorogovtsev, A.~V. Goltsev, J.~F.~F. Mendes, \href{https://link.aps.org/doi/10.1103/PhysRevE.83.051134}{Heterogeneous $k$-core versus bootstrap percolation on complex networks}, Phys. Rev. E 83 (2011) 051134.
\newblock \href {https://doi.org/10.1103/PhysRevE.83.051134} {\path{doi:10.1103/PhysRevE.83.051134}}.
\newline\urlprefix\url{https://link.aps.org/doi/10.1103/PhysRevE.83.051134}

\bibitem{Baxter_PRX_kCore}
G.~J. Baxter, S.~N. Dorogovtsev, K.-E. Lee, J.~F.~F. Mendes, A.~V. Goltsev, \href{https://link.aps.org/doi/10.1103/PhysRevX.5.031017}{Critical dynamics of the $k$-core pruning process}, Phys. Rev. X 5 (2015) 031017.
\newblock \href {https://doi.org/10.1103/PhysRevX.5.031017} {\path{doi:10.1103/PhysRevX.5.031017}}.
\newline\urlprefix\url{https://link.aps.org/doi/10.1103/PhysRevX.5.031017}

\bibitem{Kahng_2016_PRE_kCore}
D.~Lee, M.~Jo, B.~Kahng, \href{https://link.aps.org/doi/10.1103/PhysRevE.94.062307}{Critical behavior of $k$-core percolation: Numerical studies}, Phys. Rev. E 94 (2016) 062307.
\newblock \href {https://doi.org/10.1103/PhysRevE.94.062307} {\path{doi:10.1103/PhysRevE.94.062307}}.
\newline\urlprefix\url{https://link.aps.org/doi/10.1103/PhysRevE.94.062307}

\bibitem{Ahmed_2007_pqCore}
A.~Ahmed, V.~Batagelj, X.~Fu, S.-H. Hong, D.~Merrick, A.~Mrvar, Visualisation and analysis of the internet movie database, 2007 6th International Asia-Pacific Symposium on Visualization (2007) 17--24.

\bibitem{Cerinsek_2015_ijCore}
M.~Cerinsek, V.~Batagelj, \href{http://dblp.uni-trier.de/db/journals/socnet/socnet42.html#CerinsekB15}{Generalized two-mode cores.}, Soc. Networks 42 (2015) 80--87.
\newline\urlprefix\url{http://dblp.uni-trier.de/db/journals/socnet/socnet42.html#CerinsekB15}

\bibitem{Liu_2020_ijCore}
B.~Liu, L.~Yuan, X.~Lin, L.~Qin, W.~Zhang, J.~Zhou, \href{https://doi.org/10.1007/s00778-020-00606-9}{Efficient ($\alpha$, $\beta$)-core computation in bipartite graphs}, VLDB J. 29~(5) (2020) 1075--1099.
\newblock \href {https://doi.org/10.1007/s00778-020-00606-9} {\path{doi:10.1007/s00778-020-00606-9}}.
\newline\urlprefix\url{https://doi.org/10.1007/s00778-020-00606-9}

\bibitem{LaurentHD_2021_PRL_Social_Confinement}
G.~St-Onge, V.~Thibeault, A.~Allard, L.~J. Dub\'e, L.~H\'ebert-Dufresne, \href{https://link.aps.org/doi/10.1103/PhysRevLett.126.098301}{Social confinement and mesoscopic localization of epidemics on networks}, Phys. Rev. Lett. 126 (2021) 098301.
\newblock \href {https://doi.org/10.1103/PhysRevLett.126.098301} {\path{doi:10.1103/PhysRevLett.126.098301}}.
\newline\urlprefix\url{https://link.aps.org/doi/10.1103/PhysRevLett.126.098301}

\bibitem{Aksoy2020}
S.~G. Aksoy, C.~Joslyn, C.~Ortiz Marrero, B.~Praggastis, E.~Purvine, \href{https://doi.org/10.1140/epjds/s13688-020-00231-0}{Hypernetwork science via high-order hypergraph walks}, EPJ Data Science 9~(1) (2020) 16.
\newblock \href {https://doi.org/10.1140/epjds/s13688-020-00231-0} {\path{doi:10.1140/epjds/s13688-020-00231-0}}.
\newline\urlprefix\url{https://doi.org/10.1140/epjds/s13688-020-00231-0}

\bibitem{ER_1960_evolution}
P.~Erd{\H{o}}s, A.~R{\'e}nyi, et~al., \href{https://www.renyi.hu/~p_erdos/1959-11.pdf}{On the evolution of random graphs}, Publ. Math. Inst. Hung. Acad. Sci 5~(1) (1960) 17--60.
\newline\urlprefix\url{https://www.renyi.hu/~p_erdos/1959-11.pdf}

\bibitem{Newman_2001_PRE_GeneratingFunctions}
M.~E.~J. Newman, S.~H. Strogatz, D.~J. Watts, \href{https://link.aps.org/doi/10.1103/PhysRevE.64.026118}{Random graphs with arbitrary degree distributions and their applications}, Phys. Rev. E 64 (2001) 026118.
\newblock \href {https://doi.org/10.1103/PhysRevE.64.026118} {\path{doi:10.1103/PhysRevE.64.026118}}.
\newline\urlprefix\url{https://link.aps.org/doi/10.1103/PhysRevE.64.026118}

\bibitem{Watts1998}
D.~J. Watts, S.~H. Strogatz, \href{http://www.nature.com/articles/30918}{{Collective dynamics of ‘small-world' networks}}, Nature 393~(6684) (1998) 440--442.
\newblock \href {https://doi.org/10.1038/30918} {\path{doi:10.1038/30918}}.
\newline\urlprefix\url{http://www.nature.com/articles/30918}

\bibitem{Barabasi1999}
A.-L. Barab{\'{a}}si, R.~Albert, \href{https://www.sciencemag.org/lookup/doi/10.1126/science.286.5439.509 https://arxiv.org/pdf/cond-mat/9910332}{{Emergence of Scaling in Random Networks}}, Science 286~(5439) (1999) 509--512.
\newblock \href {https://doi.org/10.1126/science.286.5439.509} {\path{doi:10.1126/science.286.5439.509}}.
\newline\urlprefix\url{https://www.sciencemag.org/lookup/doi/10.1126/science.286.5439.509 https://arxiv.org/pdf/cond-mat/9910332}

\bibitem{albert_statistical_2002}
R.~Albert, A.-L. Barabási, \href{https://link.aps.org/doi/10.1103/RevModPhys.74.47}{Statistical mechanics of complex networks}, Reviews of Modern Physics 74~(1) (2002) 47--97.
\newblock \href {https://doi.org/10.1103/RevModPhys.74.47} {\path{doi:10.1103/RevModPhys.74.47}}.
\newline\urlprefix\url{https://link.aps.org/doi/10.1103/RevModPhys.74.47}

\bibitem{dorogovtsev_evolution_2002}
S.~N. Dorogovtsev, J.~F.~F. Mendes, \href{https://doi.org/10.1080/00018730110112519}{Evolution of networks}, Advances in Physics 51~(4) (2002) 1079--1187.
\newblock \href {https://doi.org/10.1080/00018730110112519} {\path{doi:10.1080/00018730110112519}}.
\newline\urlprefix\url{https://doi.org/10.1080/00018730110112519}

\bibitem{newman_structure_2003}
M.~E.~J. Newman, \href{http://epubs.siam.org/doi/10.1137/S003614450342480}{The structure and function of complex networks}, {SIAM} Review 45~(2) (2003) 167--256.
\newblock \href {https://doi.org/10.1137/S003614450342480} {\path{doi:10.1137/S003614450342480}}.
\newline\urlprefix\url{http://epubs.siam.org/doi/10.1137/S003614450342480}

\bibitem{Yongsun_2021_CHA_HPT}
Y.~Lee, J.~Lee, S.~M. Oh, D.~Lee, B.~Kahng, \href{https://doi.org/10.1063/5.0047608}{Homological percolation transitions in growing simplicial complexes}, Chaos 31~(4) (2021) 041102.
\newblock \href {https://doi.org/10.1063/5.0047608} {\path{doi:10.1063/5.0047608}}.
\newline\urlprefix\url{https://doi.org/10.1063/5.0047608}

\bibitem{Human_Genome_Project_2001_Science}
J.~C. Venter, et~al., \href{https://www.science.org/doi/abs/10.1126/science.1058040}{The sequence of the human genome}, Science 291~(5507) (2001) 1304--1351.
\newblock \href {https://doi.org/10.1126/science.1058040} {\path{doi:10.1126/science.1058040}}.
\newline\urlprefix\url{https://www.science.org/doi/abs/10.1126/science.1058040}

\bibitem{Yeast_Genetic_Interaction_2004_Science}
A.~H.~Y. Tong, et~al., \href{https://www.science.org/doi/abs/10.1126/science.1091317}{Global mapping of the yeast genetic interaction network}, Science 303~(5659) (2004) 808--813.
\newblock \href {https://doi.org/10.1126/science.1091317} {\path{doi:10.1126/science.1091317}}.
\newline\urlprefix\url{https://www.science.org/doi/abs/10.1126/science.1091317}

\bibitem{Protein_Interaction_Map_of_Drosophila_melanogaster_2003_Science}
L.~Giot, et~al., \href{https://www.science.org/doi/abs/10.1126/science.1090289}{A protein interaction map of \textit{Drosophila melanogaster}}, Science 302~(5651) (2003) 1727--1736.
\newblock \href {https://doi.org/10.1126/science.1090289} {\path{doi:10.1126/science.1090289}}.
\newline\urlprefix\url{https://www.science.org/doi/abs/10.1126/science.1090289}

\end{thebibliography}
\end{document}